\documentclass[12pt,a4paper]{article}
\usepackage{amssymb}
\usepackage{amsmath}
\usepackage{latexsym}
\usepackage{cite}
\usepackage{physics}
\begin{document}

\title{\vspace{-1cm}\bf Nonlinear corrections in the quantization\\ of a weakly nonideal Bose gas at zero\\ temperature. II. The general case}

\author{
Mikhail~N.~Smolyakov
\\
{\small{\em Skobeltsyn Institute of Nuclear Physics, Lomonosov Moscow
State University,
}}\\
{\small{\em Moscow 119991, Russia}}}

\date{}
\maketitle

\begin{abstract}
In the present paper, discussion of the canonical quantization of a weakly nonideal Bose gas at zero temperature within the framework of the Bogolyubov approach is continued. Contrary to the previous paper on this subject, here the two-body interaction potential is considered in the general form. It is shown that in such a case consideration of the first nonlinear correction also leads to the automatic particle number conservation without any additional assumptions or modification of the resulting effective Hamiltonian.
\end{abstract}

\section{Introduction}
In the renowned paper \cite{Bogolyubov}, N.N.~Bogolyubov proposed an approach to quantize a weakly nonideal Bose gas at zero temperature, which led to explanation of the superfluidity phenomenon at microscopic level. Over the following years, the method of Bogolyubov was actively studied and developed, see, for example, reviews \cite{DGPS,Review,Andersen:2003qj} and books \cite{Pitaevskii3,Pethick,Kvasnikov} for a detailed discussion of the topic.

However, the method in its original form has a well-known drawback, which is a nonconservation of particle number in the effective theory describing quasi-particles, the latter manifests itself through nondiagonal terms in the resulting quadratic effective Hamiltonian. There exist several standard methods to solve this problem (see, for example, discussion in \cite{Review}), in some of which the unphysical nondiagonal part of the effective Hamiltonian is simply included into background energy of the system. Meanwhile, one should mention some fairly new approaches, which provide the particle number conservation in more consistent ways, as well as their generalizations.

The first approach is to deal with a system with a well-defined number of particles \cite{Castin}, which is achieved by removing the terms that violate the $U(1)$ symmetry by means of a projection operator (see also \cite{Gardiner:1997yk} for a different approach leading to the same results). Later, both approaches \cite{Castin,Gardiner:1997yk} were generalized \cite{GA}, including for the cases of $n$-component systems \cite{Sorensen,MG}.

The second approach was proposed in papers \cite{Leggett1,Leggett2}, in which a special particle number conserving ansatz (the Leggett ansatz) for the ground state was used. Later, this approach was also generalized to the inhomogeneous \cite{DS1} and time-dependent \cite{DS2} cases.

An interesting unification of the approaches of \cite{Castin} and \cite{Leggett1,Leggett2} can be found in \cite{JC}, in which a special ansatz for the ground state is also used (namely, the coherent state), whereas the resulting evolution equations are naturally reduced to those of \cite{Castin} (see \cite{JTC} for a further development of the approach).

A completely different method to maintain the particle number conservation was proposed in paper \cite{Smolyakov:2021vlo}, in which it was demonstrated that the use of only the linear approximation within the Bogolyubov approach is in principle insufficient for a correct description of even free quasi-particles, and the use of only the linear approximation is exactly the origin of the resulting particle number nonconservation. Meanwhile, the use of additional nonoscillation modes in the linear approximation (these modes are exact solutions of the linearized Heisenberg equation) together with the use of the first nonlinear correction allows one to slightly modify the method of Bogolyubov and to solve the problem of nonconserved particle number in a natural way (i.e., without any modification of the initial Hamiltonian of the system or the resulting effective Hamiltonian, as well as without any additional assumptions about the properties of the ground state of the system). Although technically the approach proposed in \cite{Smolyakov:2021vlo} may look quite complicated, in essence it is very simple --- the terms $(\textit{zero order})\times(\textit{first nonlinear correction})$ are of the same order as the terms $(\textit{linear order})^{2}$, so they compensate the unphysical nondiagonal terms in the operator of particle number (and, consequently, in the effective Hamiltonian), which emerge from the terms $(\textit{linear order})^{2}$ when only the linear approximation is taken into account, and ensure the automatic particle number conservation at the quadratic level.\footnote{The same method also works for perturbations against a nontopological soliton in the nonlinear Schr\"{o}dinger equation at the classical level \cite{Smolyakov}.} Moreover, it turns out that formally the approximate solution of the Heisenberg equation discussed in \cite{Smolyakov:2021vlo} differs from the standard one of \cite{Bogolyubov}. It should be noted that the use of nonoscillation modes is a very important ingredient of the method.

In paper \cite{Smolyakov:2021vlo}, two-body interaction potential of the form \cite{Fetter}
\begin{equation}\label{interactpotent}
V(|\vec x-\vec y|)=g\delta^{(3)}(\vec x-\vec y),
\end{equation}
where $g>0$, was considered, leading to the well-known Gross-Pitaevskii differential equation \cite{Gross,Pitaevskii1}. In the present paper, an arbitrary form of the two-body interaction potential $V(|\vec x-\vec y|)$ is considered. One might think that the successful finding of a suitable nonlinear solution of the Heisenberg equation in \cite{Smolyakov:2021vlo} was a consequence of the fact that for the simplified two-body interaction potential \eqref{interactpotent} the Heisenberg equation reduces to a standard local differential equation, while in the case of arbitrary two-body interaction potential the Heisenberg equation reduces to a {\em nonlocal} integro-differential equation. However, it will be shown below that a suitable nonlinear solution of the latter equation can be found as well.

This paper is more technical than the previous paper \cite{Smolyakov:2021vlo}. A detailed discussion of the method can be found in \cite{Smolyakov:2021vlo}, so I will not reproduce it below. Moreover, some of the bulky calculations, which are necessary for calculating the canonical commutation relations, fully coincide with those already performed in \cite{Smolyakov:2021vlo}, so I will refer to \cite{Smolyakov:2021vlo} when necessary. The main notations used in the present paper fully coincide with those used in \cite{Smolyakov:2021vlo}.

Basically, the present paper has the structure that is similar to the one of \cite{Smolyakov:2021vlo}. It is organized as follows. In Section~2 the basic setup is described. In Section~3 equations of motion and commutation relations for perturbations are derived. In Section~4 the linear approximation, including the nonoscillation modes, is discussed. In Section~5 an explicit solution for the first nonlinear correction is presented and the corresponding commutation relations are shown to be satisfied. In Section~6 operators of the resulting integrals of motion are calculated. The obtained results are briefly discussed in Conclusion. Appendix contains an auxiliary material.

\section{Setup}
As in paper \cite{Bogolyubov}, within the second quantization formalism we take the standard Hamiltonian
\begin{equation}\label{initHamiltonian}
\hat H=\int d^{3}x\left(\frac{\hbar^{2}}{2m}\sum\limits_{i=1}^{3}\partial_{i}\hat\Psi^{\dagger}(t,\vec x)\partial_{i}\hat\Psi(t,\vec x)+\frac{1}{2}\int d^{3}y\hat\Psi^{\dagger}(t,\vec x)\hat\Psi^{\dagger}(t,\vec y)V(|\vec x-\vec y|)\hat\Psi(t,\vec x)\hat\Psi(t,\vec y)\right)
\end{equation}
for a system in a spatial ``box'' of size $L\times L\times L$ with periodic boundary conditions for $\hat\Psi(t,\vec x)$. The operators $\hat\Psi(t,\vec x)$ and $\hat\Psi^{\dagger}(t,\vec x)$ are supposed to satisfy the standard canonical commutation relations
\begin{equation}\label{commrel}
[\hat\Psi(t,\vec x),\hat\Psi^{\dagger}(t,\vec y)]=\delta^{(3)}(\vec x-\vec y),\qquad
[\hat\Psi(t,\vec x),\hat\Psi(t,\vec y)]=0.
\end{equation}
The operator of the conserved particle number is
\begin{equation}
\hat N=\int d^{3}x\hat\Psi^{\dagger}(t,\vec x)\hat\Psi(t,\vec x).
\end{equation}
For the Hamiltonian $\hat H$ defined by \eqref{initHamiltonian} and with \eqref{commrel}, the Heisenberg equation
\begin{equation}\label{Heiseq}
\frac{d\hat\Psi(t,\vec x)}{dt}=\frac{i}{\hbar}[\hat H,\hat\Psi(t,\vec x)]
\end{equation}
leads to the well-known nonlinear integro-differential equation for the operator $\hat\Psi(t,\vec x)$:
\begin{equation}\label{eqmotgeneral}
i\hbar\dot{\hat\Psi}(t,\vec x)=-\frac{\hbar^{2}}{2m}\Delta\hat\Psi(t,\vec x)+\int d^{3}yV(|\vec x-\vec y|)\hat\Psi^{\dagger}(t,\vec y)\hat\Psi(t,\vec y)\hat\Psi(t,\vec x),
\end{equation}
where $\dot{}=\frac{\partial}{\partial t}$ and $\Delta=\sum\limits_{i=1}^{3}\partial_{i}^{2}$.

Since the occupation number is supposed to be large, the creation and annihilation operators of condensed particles can be replaced by $c$-numbers \cite{Bogolyubov}. This replacement corresponds to a classical solution of Eq.~\eqref{eqmotgeneral} of the form
\begin{equation}\label{backgrsol0}
\Psi_{0}(t)=\sqrt{\frac{\omega}{g}}e^{-\frac{i}{\hbar}\omega t}
\end{equation}
with $\omega>0$ and
\begin{equation}
g=\int d^{3}yV(|\vec x-\vec y|)>0.
\end{equation}
For solution \eqref{backgrsol0}, its energy and particle number are
\begin{equation}\label{backgrEN}
E_{0}=\frac{\omega^{2}L^{3}}{2\,g},\quad N_{0}=\frac{\omega L^{3}}{g}.
\end{equation}

\section{Equations of motion for perturbations, integrals of motion and commutation relations}
Let us represent the operator $\hat\Psi(t,\vec x)$ as
\begin{equation}\label{representInit}
\hat\Psi(t,\vec x)=e^{-\frac{i}{\hbar}\omega t}\left(\sqrt{\frac{\omega}{g}}+\hat\psi(t,\vec x)\right),
\end{equation}
where $\hat\psi(t,\vec x)$ satisfies the full nonlinear integro-differential equation of motion, which follows directly from Eq.~\eqref{eqmotgeneral}:
\begin{align}\nonumber
&i\hbar\dot{\hat\psi}(t,\vec x)=-\frac{\hbar^{2}}{2m}\Delta\hat\psi(t,\vec x)+\frac{\omega}{g}\int d^{3}yV(|\vec x-\vec y|)\left(\hat\psi(t,\vec y)+\hat\psi^{\dagger}(t,\vec y)\right)\\\label{eqnlcquant}
&+\sqrt{\frac{\omega}{g}}\int d^{3}yV(|\vec x-\vec y|)\left[\left(\hat\psi(t,\vec y)+\hat\psi^{\dagger}(t,\vec y)\right)\hat\psi(t,\vec x)
+\hat\psi^{\dagger}(t,\vec y)\hat\psi(t,\vec y)\right]+...\,\,.
\end{align}
Here ``$...$'' denotes the terms of the higher orders in $\hat\psi^{\dagger}(t,\vec x)$ and $\hat\psi(t,\vec x)$. In order to solve this equation using the methods of perturbation theory, we can expand the operator $\hat\psi(t,\vec x)$ into the series
\begin{equation}\label{seriesPhiinf}
\hat\psi(t,\vec x)=\hat\varphi(t,\vec x)+\sum\limits_{n=1}^{\infty}\beta^{n}\hat\phi_{n}(t,\vec x),
\end{equation}
where $\beta=\frac{1}{\sqrt{N_{0}}}=\sqrt{\frac{g}{\omega L^{3}}}\ll 1$ and $\hat\varphi(t,\vec x)$ is a solution of the well-known linear integro-differential equation
\begin{equation}\label{lineq}
i\hbar\dot{\hat\varphi}(t,\vec x)=-\frac{\hbar^{2}}{2m}\Delta\hat\varphi(t,\vec x)+\frac{\omega}{g}\int d^{3}yV(|\vec x-\vec y|)\left(\hat\varphi(t,\vec y)+\hat\varphi^{\dagger}(t,\vec y)\right),
\end{equation}
which follows from Eq.~\eqref{eqnlcquant}. Of course, analytical finding of all terms $\hat\phi_{n}(t,\vec x)$ in \eqref{seriesPhiinf} is impossible, so, as in \cite{Smolyakov:2021vlo}, we will restrict ourselves to the linear and quadratic orders only. Let us denote the first nonlinear correction $\hat\phi_{1}(t,\vec x)$ as $\hat\phi(t,\vec x)$, reduce the operator $\hat\psi(t,\vec x)$ to
\begin{equation}\label{representNL}
\hat\psi(t,\vec x)=\hat\varphi(t,\vec x)+\beta\hat\phi(t,\vec x),
\end{equation}
and substitute this simplified representation into Eq.~\eqref{eqnlcquant}. Combining all terms of the order of $\beta$, we obtain the inhomogeneous integro-differential equation
\begin{align}\nonumber
&i\hbar\dot{\hat\phi}(t,\vec x)+\frac{\hbar^{2}}{2m}\Delta\hat\phi(t,\vec x)-\frac{\omega}{g}\int d^{3}yV(|\vec x-\vec y|)\left(\hat\phi(t,\vec y)+\hat\phi^{\dagger}(t,\vec y)\right)\\\label{eqnlcquant2}
&=\frac{\omega\sqrt{L^{3}}}{g}\int d^{3}yV(|\vec x-\vec y|)\Bigl(\left(\hat\varphi(t,\vec y)+\hat\varphi^{\dagger}(t,\vec y)\right)\hat\varphi(t,\vec x)
+\hat\varphi^{\dagger}(t,\vec y)\hat\varphi(t,\vec y)\Bigr).
\end{align}
In principle, the nonlinear correction $\hat\phi(t,\vec x)$ can be considered as a backreaction. In this connection, it should be noted that the ansatz fully analogous to \eqref{representNL} was used in \cite{Schutzhold:2005ex} for examining backreaction of linear quantum perturbations on the classical background of the system.\footnote{In \cite{Schutzhold:2005ex}, an analogy with quantum field theory in curved spacetime was used for the analysis. See also \cite{Baak:2022hum} for a further development of the perturbative approach of \cite{Schutzhold:2005ex}.}

It was shown in \cite{Smolyakov:2021vlo} that for representation \eqref{representInit} the operator of particle number of perturbations looks like (this is the exact result)
\begin{equation}\label{Npquadrpsi}
\hat N_{p}=\hat N-N_{0}=\int d^{3}x\left(\sqrt{\frac{\omega}{g}}\left(\hat\psi^{\dagger}+\hat\psi\right)+\hat\psi^{\dagger}\hat\psi\right).
\end{equation}
Substituting representation \eqref{representInit} into \eqref{initHamiltonian}, eliminating the terms with $\partial_{i}$ by means of the nonlinear Eq.~\eqref{eqnlcquant}, and retaining only linear and quadratic in $\hat\psi$ and $\hat\psi^{\dagger}$ terms, we obtain
\begin{equation}\label{Epquadrpsi}
\hat E_{p}=\hat E-E_{0}=\omega\hat N_{p}+\frac{i\hbar}{2}\int d^{3}x\left(\hat\psi^{\dagger}\dot{\hat\psi}-\dot{\hat\psi}^{\dagger}\hat\psi\right).
\end{equation}
Although Eq.~\eqref{eqnlcquant} differs from the analogous equation in \cite{Smolyakov:2021vlo}, operator \eqref{Epquadrpsi} coincides with the analogous operator in \cite{Smolyakov:2021vlo}. Then, substituting representation \eqref{representNL} into \eqref{Npquadrpsi} and \eqref{Epquadrpsi}, using the fact that $\sqrt{\frac{\omega}{g}}=\frac{1}{\beta\sqrt{L^{3}}}$, and retaining only the terms of the order of $\beta^{-1}$ and $\beta^{0}=1$, we get
\begin{eqnarray}\label{NpquadrQNL}
&&\hat N_{p}=\int d^{3}x\left(\sqrt{\frac{\omega}{g}}\left(\hat\varphi^{\dagger}+\hat\varphi\right)+\hat\varphi^{\dagger}\hat\varphi
+\frac{1}{\sqrt{L^{3}}}\left(\hat\phi^{\dagger}+\hat\phi\right)\right),
\\\label{EpquadrQNL}
&&\hat E_{p}=\omega\hat N_{p}+\frac{i\hbar}{2}\int d^{3}x\left(\hat\varphi^{\dagger}\dot{\hat\varphi}-\dot{\hat\varphi}^{\dagger}\hat\varphi\right),
\end{eqnarray}
which have exactly the same forms as the corresponding operators in \cite{Smolyakov:2021vlo}. It is exactly the term $\sim\left(\hat\phi^{\dagger}+\hat\phi\right)$ that provides contributions compensating the unphysical nondiagonal operators in \eqref{NpquadrQNL} originating from the term $\sim\hat\varphi^{\dagger}\hat\varphi$.

Now let us consider the canonical commutation relations. Suppose that for the linear solution the commutation relations
\begin{eqnarray}\label{CCR01}
&&[\hat\varphi(t,\vec x),\hat\varphi(t,\vec y)]=0,\\\label{CCR02}
&&[\hat\varphi(t,\vec x),\hat\varphi^{\dagger}(t,\vec y)]=\delta^{(3)}(\vec x-\vec y)
\end{eqnarray}
are exactly satisfied. Since in our approximation only the first nonlinear correction is taken into account, with \eqref{CCR01} and \eqref{CCR02} the relevant parts of the canonical commutation relations are reduced to \cite{Smolyakov:2021vlo}
\begin{align}
&[\hat\psi(t,\vec x),\hat\psi(t,\vec y)]\to 0+\beta[\hat\phi(t,\vec x),\hat\varphi(t,\vec y)]+\beta[\hat\varphi(t,\vec x),\hat\phi(t,\vec y)]=0,\\
&[\hat\psi(t,\vec x),\hat\psi^{\dagger}(t,\vec y)]\to\delta^{(3)}(\vec x-\vec y)+\beta[\hat\phi(t,\vec x),\hat\varphi^{\dagger}(t,\vec y)]+\beta[\hat\varphi(t,\vec x),\hat\phi^{\dagger}(t,\vec y)]=\delta^{(3)}(\vec x-\vec y),
\end{align}
resulting in
\begin{align}\label{CCR1}
&[\hat\phi(t,\vec x),\hat\varphi(t,\vec y)]+[\hat\varphi(t,\vec x),\hat\phi(t,\vec y)]=0,\\\label{CCR2}
&[\hat\phi(t,\vec x),\hat\varphi^{\dagger}(t,\vec y)]+[\hat\varphi(t,\vec x),\hat\phi^{\dagger}(t,\vec y)]=0.
\end{align}
Commutation relations \eqref{CCR1} and \eqref{CCR2} imply that the canonical commutation relations are satisfied up to and including the terms $\sim\beta$.

\section{Linear approximation}
First, let us consider Eq.~\eqref{lineq}. Using the standard representation
\begin{equation}\label{Vdecomp}
V(|\vec x-\vec y|)=\frac{1}{L^{3}}\sum\limits_{j}e^{\frac{i}{\hbar}\vec k_{j}(\vec x-\vec y)}v_{j},
\end{equation}
with
\begin{equation}\label{kdef}
\vec k_{j}=\frac{2\pi\hbar}{L}\vec j,\qquad \vec j=(j_{1},j_{2},j_{3}),
\end{equation}
where $j_{1}$, $j_{2}$, $j_{3}$ are integers,\footnote{Although $\vec j$ is a vector, for simplicity the mark of a vector is omitted for the subscript ``$j$''.} $v_{-j}=v_{j}$ (follows from $V(|\vec x-\vec y|)=V(|\vec y-\vec x|)$), $v_{-j}=v^{*}_{j}$ (follows from $V^{*}(|\vec x-\vec y|)=V(|\vec x-\vec y|)$), and $v_{0}=g$, one can get the well-known solution of Eq.~\eqref{lineq}
\begin{align}\nonumber
\hat\varphi_{\textrm{o}}(t,\vec x)&=
\frac{1}{\sqrt{L^{3}}}\sum\limits_{j\neq 0}\left(c_{j}e^{-\frac{i}{\hbar}(\gamma_{j}t-\vec k_{j}\vec x)}\hat a_{j}-d_{j}e^{\frac{i}{\hbar}(\gamma_{j}t-\vec k_{j}\vec x)}\hat a^{\dagger}_{j}\right)\\\label{linsol}
&=\frac{1}{\sqrt{L^{3}}}\sum\limits_{j\neq 0}e^{\frac{i}{\hbar}\vec k_{j}\vec x}\left(c_{j}e^{-\frac{i}{\hbar}\gamma_{j}t}\hat a_{j}-d_{j}e^{\frac{i}{\hbar}\gamma_{j}t}\hat a^{\dagger}_{-j}\right)
\end{align}
with
\begin{equation}\label{cddef}
c_{j}=\frac{\omega\tilde v_{j}}{\sqrt{2\gamma_{j}\left(\frac{{\vec k_{j}}^{2}}{2m}+\omega\tilde v_{j}-\gamma_{j}\right)}},\qquad d_{j}=\sqrt{\frac{\frac{{\vec k_{j}}^{2}}{2m}+\omega\tilde v_{j}-\gamma_{j}}{2\gamma_{j}}},
\end{equation}
where $\tilde v_{j}=\frac{v_{j}}{v_{0}}=\frac{v_{j}}{g}$ and $j_{1}^2+j_{2}^2+j_{3}^2\neq 0$,
\begin{equation}\label{gammadef}
\gamma_{j}=\sqrt{
\frac{{\vec k_{j}}^{2}}{2m}\left(\frac{{\vec k_{j}}^{2}}{2m}+2\omega\tilde v_{j}\right)},
\end{equation}
and
\begin{equation}
[\hat a_{j}^{},\hat a_{l}^{\dagger}]=\delta_{jl},\qquad [\hat a_{j}^{},\hat a_{l}^{}]=0.
\end{equation}
It is easy to check that $c_{j}^{2}-d_{j}^{2}=1$. Here the operators $\hat a_{j}^{}$ and $\hat a^{\dagger}_{j}$ are annihilation and creation operators of quasi-particles.

Linear solution \eqref{linsol} as it is does not satisfy the necessary commutation relation \eqref{CCR02}, which is the consequence of the absence of the operators $\hat a_{0}$, $\hat a^{\dagger}_{0}$ in the linear solution \eqref{linsol}:
\begin{equation}\label{CCRviolat}
[\hat\varphi_{\textrm{o}}^{}(t,\vec x),\hat\varphi_{\textrm{o}}^{\dagger}(t,\vec y)]=\delta^{(3)}(\vec x-\vec y)-\frac{1}{L^{3}}.
\end{equation}
As in \cite{Smolyakov:2021vlo}, this problem can be solved by recalling that apart from the standard oscillation modes presented in the linear solution \eqref{linsol}, Eq.~\eqref{lineq} provides two other solutions, which also satisfy the periodic boundary conditions. These solutions do not depend on the spatial coordinate $\vec x$, so Eq.~\eqref{lineq} takes the form
\begin{equation}\label{lineq-nonosc}
i\hbar\dot{\hat\varphi}_{\textrm{no}}^{}(t,\vec x)=\omega\left(\hat\varphi_{\textrm{no}}^{}(t,\vec x)+\hat\varphi_{\textrm{no}}^{\dagger}(t,\vec x)\right)
\end{equation}
for these modes. This is exactly the linear equation for nonoscillation modes that emerges in the theory with potential \eqref{interactpotent}, and its solutions were found in \cite{Smolyakov:2021vlo}:
\begin{equation}\label{quantmode0}
\hat\varphi_{\textrm{no\,1}}^{}(t,\vec x)=\frac{1}{2q\sqrt{L^{3}}}\left(\hat a_{0}^{}-\hat a_{0}^{\dagger}\right)
\end{equation}
(in the classical case this Goldstone mode corresponds to the global $U(1)$ symmetry of the theory),
\begin{equation}\label{quantmode}
\hat\varphi_{\textrm{no\,2}}^{}(t,\vec x)=\frac{q}{\sqrt{L^{3}}}\left(\frac{1}{2}-\frac{i}{\hbar}\omega t\right)\left(\hat a_{0}^{}+\hat a_{0}^{\dagger}\right)
\end{equation}
(in the classical case this mode corresponds to a change of the frequency $\omega$ of the background solution \eqref{backgrsol0}, in fact it is also a Goldstone-like mode resulting from the symmetry breaking associated with a choice of the frequency $\omega$), see \cite{Smolyakov:2021vlo} for a more detailed discussion of these nonoscillation modes,\footnote{See also a discussion of the ``missing eigenvector'' in \cite{Castin}, as well as papers \cite{ZM1,ZM2} for other discussions of such modes in the case of Bose-Einstein condensates.} including an explanation that the linear growth of \eqref{quantmode} in $t$ is not dangerous and does not lead to an instability. Here $q\neq 0$ is an arbitrary dimensionless real constant and
\begin{equation}
[\hat a_{0}^{},\hat a^{\dagger}_{0}]=1,\quad [\hat a_{0}^{},\hat a_{j}^{}]=0,\quad [\hat a_{0}^{},\hat a_{j}^{\dagger}]=0,\quad j\neq 0.
\end{equation}
Now, instead of \eqref{linsol}, let us take
\begin{align}\nonumber
\hat\varphi(t,\vec x)&=
\frac{1}{\sqrt{L^{3}}}\sum\limits_{j\neq 0}\left(c_{j}e^{-\frac{i}{\hbar}(\gamma_{j}t-\vec k_{j}\vec x)}\hat a_{j}^{}-d_{j}e^{\frac{i}{\hbar}(\gamma_{j}t-\vec k_{j}\vec x)}\hat a_{j}^{\dagger}\right)\\\label{linsol2}&+\frac{1}{2q\sqrt{L^{3}}}\left(\hat a_{0}^{}-\hat a_{0}^{\dagger}\right)+\frac{q}{\sqrt{L^{3}}}\left(\frac{1}{2}-\frac{i}{\hbar}\omega t\right)\left(\hat a_{0}^{}+\hat a_{0}^{\dagger}\right).
\end{align}
It is clear that solution \eqref{linsol2} satisfies the linear Eq.~\eqref{lineq}. It is not difficult to check that \eqref{linsol2} exactly satisfies commutation relations \eqref{CCR01} and \eqref{CCR02}. The fact that nonoscillation modes recover commutation relation \eqref{CCR02} is not surprising --- these modes add to \eqref{linsol} the solutions that do not depend on the spatial coordinate, thus making the corresponding set of eigenfunctions (the one that forms the delta function in \eqref{CCR02}) complete. Analogous role is played by nonoscillation modes in theories with solitons in the nonlinear Schr\"{o}dinger equation (in particular, in the case of one-dimensional solitons there are two more nonoscillation modes --- the translational mode and the mode corresponding to the Galilean transformation) --- they also make the set of the corresponding eigenfunctions complete \cite{Yan1998}, thus allowing one to perform a consistent procedure of quantization \cite{Kovtun:2021rcm}.

\section{Second order}
\subsection{Separation of the equations of motion}
As in \cite{Smolyakov:2021vlo}, the operator $\hat\varphi(t,\vec x)$ can be represented as
\begin{equation}
\hat\varphi(t,\vec x)=\hat\varphi_{\textrm{no}}(t)+\hat\varphi_{\textrm{o}}(t,\vec x),
\end{equation}
where
\begin{equation}\label{linsolquant}
\hat\varphi_{\textrm{o}}(t,\vec x)=\sum\limits_{j\neq 0}\hat\varphi_{j}(t,\vec x)=\sum\limits_{j\neq 0}\hat{\tilde\varphi}_{j}(t,\vec x)
\end{equation}
with
\begin{align}\label{philinnodef}
&\hat\varphi_{\textrm{no}}(t)=\frac{1}{2q\sqrt{L^{3}}}\left(\hat a_{0}^{}-\hat a_{0}^{\dagger}\right)+\frac{q}{\sqrt{L^{3}}}\left(\frac{1}{2}-\frac{i}{\hbar}\omega t\right)\left(\hat a_{0}^{}+\hat a_{0}^{\dagger}\right),\\
&\hat\varphi_{j}(t,\vec x)=\frac{1}{\sqrt{L^{3}}}\left(c_{j}e^{-\frac{i}{\hbar}(\gamma_{j}t-\vec k_{j}\vec x)}\hat a_{j}^{}-d_{j}e^{\frac{i}{\hbar}(\gamma_{j}t-\vec k_{j}\vec x)}\hat a^{\dagger}_{j}\right),\\\label{varphiodecomp2}
&\hat{\tilde\varphi}_{j}(t,\vec x)=\frac{1}{\sqrt{L^{3}}}e^{\frac{i}{\hbar}\vec k_{j}\vec x}\left(c_{j}e^{-\frac{i}{\hbar}\gamma_{j}t}\hat a_{j}^{}-d_{j}e^{\frac{i}{\hbar}\gamma_{j}t}\hat a^{\dagger}_{-j}\right),
\end{align}
see \eqref{linsol}. Although both representations in \eqref{linsolquant} lead to the same results, in some cases the first representation is more useful, whereas in some cases the second representation is more useful.

Now, again as in \cite{Smolyakov:2021vlo}, we represent the operator $\hat\phi(t,\vec x)$ as
\begin{equation}\label{phidecomp}
\hat\phi(t,\vec x)=\hat\phi_{\textrm{no}}(t)+\hat\phi_{\times}(t,\vec x)+\hat\phi_{\textrm{o}}(t,\vec x).
\end{equation}
The operators $\hat\phi_{\textrm{no}}(t)$, $\hat\phi_{\times}(t,\vec x)$ and $\hat\phi_{\textrm{o}}(t,\vec x)$ in \eqref{phidecomp} are supposed to be solutions of the equations
\begin{align}\label{eqquant1}
&i\hbar\dot{\hat\phi}_{\textrm{no}}^{}(t)-\omega\left(\hat\phi_{\textrm{no}}^{}(t)+\hat\phi_{\textrm{no}}^{\dagger}(t)\right)
=\omega\sqrt{L^{3}}\left(2\hat\varphi_{\textrm{no}}^{\dagger}(t)\hat\varphi_{\textrm{no}}^{}(t)+\hat\varphi_{\textrm{no}}^{2}(t)\right),\\
\nonumber
&i\hbar\dot{\hat\phi}_{\times}^{}(t,\vec x)+\frac{\hbar^{2}}{2m}\Delta\hat\phi_{\times}^{}(t,\vec x)-\frac{\omega}{g}\int d^{3}yV(|\vec x-\vec y|)\left(\hat\phi_{\times}^{}(t,\vec y)+\hat\phi_{\times}^{\dagger}(t,\vec y)\right)\\\label{eqquant2}
&=\frac{\omega\sqrt{L^{3}}}{g}\int d^{3}yV(|\vec x-\vec y|)\Bigl(\left(\hat\varphi_{\textrm{no}}^{}(t)+\hat\varphi_{\textrm{no}}^{\dagger}(t)\right)\left(\hat\varphi_{\textrm{o}}^{}(t,\vec y)+\hat\varphi_{\textrm{o}}^{}(t,\vec x)\right)+2\hat\varphi_{\textrm{no}}^{}(t)\hat\varphi_{\textrm{o}}^{\dagger}(t,\vec y)\Bigr),\\\nonumber
&i\hbar\dot{\hat\phi}_{\textrm{o}}^{}(t,\vec x)+\frac{\hbar^{2}}{2m}\Delta\hat\phi_{\textrm{o}}^{}(t,\vec x)-\frac{\omega}{g}\int d^{3}yV(|\vec x-\vec y|)\left(\hat\phi_{\textrm{o}}^{}(t,\vec y)+\hat\phi_{\textrm{o}}^{\dagger}(t,\vec y)\right)\\\label{eqquant3}
&=\frac{\omega\sqrt{L^{3}}}{g}\int d^{3}yV(|\vec x-\vec y|)\Bigl(\left(\hat\varphi_{\textrm{o}}^{}(t,\vec y)+\hat\varphi_{\textrm{o}}^{\dagger}(t,\vec y)\right)\hat\varphi_{\textrm{o}}^{}(t,\vec x)+\hat\varphi_{\textrm{o}}^{\dagger}(t,\vec y)\hat\varphi_{\textrm{o}}^{}(t,\vec y)\Bigr),
\end{align}
which follow from Eq.~\eqref{eqnlcquant2}.

\subsection{Operator $\hat\phi_{\textrm{no}}^{}(t)$ for the nonoscillation modes}
Eq.~\eqref{eqquant1} fully coincides with the analogous equation in \cite{Smolyakov:2021vlo}, which in the explicit form looks like
\begin{align}\nonumber
i\hbar\dot{\hat\phi}_{\textrm{no}}^{}-\omega\left(\hat\phi_{\textrm{no}}^{}+\hat\phi_{\textrm{no}}^{\dagger}\right)
&=\frac{\omega}{\sqrt{L^{3}}}\left(
-\frac{1}{4q^{2}}\left(\hat a_{0}^{}-\hat a_{0}^{\dagger}\right)^{2}+\left(\frac{1}{2}+\frac{i\omega t}{\hbar}\right)\left(\hat a_{0}^{2}-\hat a_{0}^{\dagger\,2}\right)\right.\\\label{eqquant1expl}
&+\left. q^{2}\left(\frac{3}{4}-\frac{i\omega t}{\hbar}+\frac{\omega^{2}t^{2}}{\hbar^{2}}\right)\left(\hat a_{0}^{}+\hat a_{0}^{\dagger}\right)^{2}-1\right).
\end{align}
Its solution was found in \cite{Smolyakov:2021vlo} and has the form
\begin{align}\nonumber
\hat\phi_{\textrm{no}}(t)&=\frac{\omega}{2\sqrt{L^{3}}}\left(
\frac{1}{4\omega q^{2}}\left(\hat a_{0}^{}-\hat a_{0}^{\dagger}\right)^{2}+\left(\frac{1}{2\omega}-\frac{it}{\hbar}\right)\left(\hat a_{0}^{2}-\hat a_{0}^{\dagger\,2}\right)\right.\\
\label{phinosolfull}&\left.-q^{2}\left(\frac{1}{4\omega}+\frac{it}{\hbar}+\frac{\omega t^{2}}{\hbar^{2}}\right)\left(\hat a_{0}^{}+\hat a_{0}^{\dagger}\right)^{2}+\frac{1}{\omega}\right).
\end{align}
It was shown in \cite{Smolyakov:2021vlo} that \eqref{phinosolfull} satisfies the necessary commutation relations \eqref{CCR1} and \eqref{CCR2}.

\subsection{Cross term $\hat\phi_{\times}(t,\vec x)$}
Using representation \eqref{Vdecomp}, one can check that
\begin{equation}\label{Vexp}
\int d^{3}yV(|\vec x-\vec y|)e^{\frac{i}{\hbar}\vec k_{j}\vec y}=e^{\frac{i}{\hbar}\vec k_{j}\vec x}v_{j}.
\end{equation}
Using the second representation for $\hat\varphi_{\textrm{o}}(t,\vec x)$ in \eqref{linsolquant} with \eqref{varphiodecomp2} and applying \eqref{Vexp}, after straightforward calculations Eq.~\eqref{eqquant2} can be brought to the form
\begin{align}\nonumber
&i\hbar\dot{\hat\phi}_{\times}^{}(t,\vec x)+\frac{\hbar^{2}}{2m}\Delta\hat\phi_{\times}^{}(t,\vec x)-\frac{\omega}{g}\int d^{3}yV(|\vec x-\vec y|)\left(\hat\phi_{\times}^{}(t,\vec y)+\hat\phi_{\times}^{\dagger}(t,\vec y)\right)=\frac{\omega}{\sqrt{L^{3}}}\sum\limits_{j\neq 0}e^{\frac{i}{\hbar}\vec k_{j}\vec x}\\\nonumber
&\times\left[e^{-\frac{i}{\hbar}\gamma_{j}t}\hat a_{j}\left(q\left((1+\tilde v_{j})c_{j}-\tilde v_{j}d_{j}\right)\left(\hat a_{0}^{}+\hat a_{0}^{\dagger}\right)-\frac{1}{q}\tilde v_{j}d_{j}\left(\hat a_{0}^{}-\hat a_{0}^{\dagger}\right)+tq\frac{2i\omega}{\hbar}\tilde v_{j}d_{j}\left(\hat a_{0}^{}+\hat a_{0}^{\dagger}\right)\right)
\right.\\\label{eqquant2expl}&\left.
+\,e^{\frac{i}{\hbar}\gamma_{j}t}\hat a_{-j}^{\dagger}\left(q(\tilde v_{j}c_{j}-(1+\tilde v_{j})d_{j})\left(\hat a_{0}^{}+\hat a_{0}^{\dagger}\right)+\frac{1}{q}\tilde v_{j}c_{j}\left(\hat a_{0}^{}-\hat a_{0}^{\dagger}\right)-tq\frac{2i\omega}{\hbar}\tilde v_{j}c_{j}\left(\hat a_{0}^{}+\hat a_{0}^{\dagger}\right)\right)
\right].
\end{align}
The terms $\sim t\,e^{\mp\frac{i}{\hbar}\gamma_{j}t}$ in Eq.~\eqref{eqquant2expl} imply a solution of Eq.~\eqref{eqquant2expl} of the form \cite{Smolyakov:2021vlo}
\begin{align}\nonumber
\hat\phi_{\times}^{}(t,\vec x)&=\frac{\omega}{\sqrt{L^{3}}}\sum\limits_{j\neq 0}e^{\frac{i}{\hbar}\vec k_{j}\vec x}\left(e^{-\frac{i}{\hbar}\gamma_{j}t}\left(\hat A_{j}+t\hat B_{j}\right)+e^{\frac{i}{\hbar}\gamma_{j}t}\left(\hat C_{j}+t\hat D_{j}\right)\right)\\\label{substX}
&+\frac{\omega}{\sqrt{L^{3}}}\sum\limits_{j\neq 0}e^{\frac{i}{\hbar}\vec k_{j}\vec x}\left(e^{-\frac{i}{\hbar}\gamma_{j}t}c_{j}\hat Q_{j}-e^{\frac{i}{\hbar}\gamma_{j}t}d_{j}\hat Q_{-j}^{\dagger}\right).
\end{align}
Here the second sum is just a solution of the homogeneous part of Eq.~\eqref{eqquant2expl}, which can always be added to a particular solution of the inhomogeneous equation. Substituting \eqref{substX} into \eqref{eqquant2expl} and combining the similar terms, we get the algebraic system of equations
\begin{align}\nonumber
&\left(\gamma_{j}-\frac{{\vec k_{j}}^{2}}{2m}-\omega\tilde v_{j}\right)\hat A_{j}+i\hbar\hat B_{j}-\omega\tilde v_{j}\hat C_{-j}^{\dagger}\\&=q\bigl((1+\tilde v_{j})c_{j}-\tilde v_{j}d_{j}\bigr)\left(\hat a_{0}^{}+\hat a_{0}^{\dagger}\right)\hat a_{j}-\frac{1}{q}\tilde v_{j}d_{j}\left(\hat a_{0}^{}-\hat a_{0}^{\dagger}\right)\hat a_{j},\\\nonumber
&\left(-\gamma_{j}-\frac{{\vec k_{j}}^{2}}{2m}-\omega\tilde v_{j}\right)\hat C_{j}+i\hbar\hat D_{j}-\omega\tilde v_{j}\hat A_{-j}^{\dagger}\\&=q\bigl(\tilde v_{j}c_{j}-(1+\tilde v_{j})d_{j}\bigr)\left(\hat a_{0}^{}+\hat a_{0}^{\dagger}\right)\hat a_{-j}^{\dagger}+\frac{1}{q}\tilde v_{j}c_{j}\left(\hat a_{0}^{}-\hat a_{0}^{\dagger}\right)\hat a_{-j}^{\dagger},\\
&\left(\gamma_{j}-\frac{{\vec k_{j}}^{2}}{2m}-\omega\tilde v_{j}\right)\hat B_{j}-\omega\tilde v_{j}\hat D_{-j}^{\dagger}=\frac{2i\omega}{\hbar}q\tilde v_{j}d_{j}\left(\hat a_{0}^{}+\hat a_{0}^{\dagger}\right)\hat a_{j},\\
&\left(-\gamma_{j}-\frac{{\vec k_{j}}^{2}}{2m}-\omega\tilde v_{j}\right)\hat D_{j}-\omega\tilde v_{j}\hat B_{-j}^{\dagger}=-\frac{2i\omega}{\hbar}q\tilde v_{j}c_{j}\left(\hat a_{0}^{}+\hat a_{0}^{\dagger}\right)\hat a_{-j}^{\dagger}.
\end{align}
Let us take the following solution of this system of equations:
\begin{align}\label{Xcoeff1}
&\hat A_{j}=\frac{1}{q\omega}c_{j}\left(\hat a_{0}^{}-\hat a_{0}^{\dagger}\right)\hat a_{j}^{},\\\label{Xcoeff2}
&\hat B_{j}=-\frac{iq}{\hbar}\left(1+\frac{{\vec k_{j}}^{2}\tilde v_{j}}{2m\gamma_{j}}\right)c_{j}\left(\hat a_{0}^{}+\hat a_{0}^{\dagger}\right)\hat a_{j}^{},\\\label{Xcoeff3}
&\hat C_{j}=\frac{q}{\omega^{2}\tilde v_{j}}\left(\gamma_{j}-\frac{{\vec k_{j}}^{2}}{2m}-\omega\tilde v_{j}\right)\frac{{\vec k_{j}}^{2}}{2m\gamma_{j}}\,c_{j}\left(\hat a_{0}^{}+\hat a_{0}^{\dagger}\right)\hat a_{-j}^{\dagger},\\\label{Xcoeff4}
&\hat D_{j}=\frac{iq}{\hbar}\left(1-\frac{\gamma_{j}(1+\tilde v_{j})}{\omega\tilde v_{j}}+\frac{{\vec k_{j}}^{2}}{2m\gamma_{j}}\left(\tilde v_{j}+\frac{\gamma_{j}(1+\tilde v_{j})}{\omega\tilde v_{j}}\right)\right)c_{j}\left(\hat a_{0}^{}+\hat a_{0}^{\dagger}\right)\hat a_{-j}^{\dagger}.
\end{align}
In addition, let us take
\begin{equation}\label{deltaphiXQ}
\hat Q_{j}=\frac{q d_{j}^{2}}{\omega}\frac{{\vec k_{j}}^{2}}{2m\gamma_{j}}\left(\hat a_{0}^{}+\hat a_{0}^{\dagger}\right)\hat a_{j}+\frac{1}{2q\omega}\left(\frac{{\vec k_{j}}^{2}\tilde v_{j}}{2m\gamma_{j}}-1\right)\left(\hat a_{0}^{}-\hat a_{0}^{\dagger}\right)\hat a_{j},
\end{equation}
the necessity for the extra term with $\hat Q_{j}$ is explained in \cite{Smolyakov:2021vlo}.

By performing straightforward calculations, one can check that for solution \eqref{substX} with \eqref{Xcoeff1}--\eqref{deltaphiXQ} the necessary commutation relations take the form
\begin{align}\nonumber
&[\hat\phi_{\times}(t,\vec x),\hat\varphi(t,\vec y)]+[\hat\varphi(t,\vec x),\hat\phi_{\times}(t,\vec y)]=\frac{\omega}{L^{3}}\sum\limits_{j\neq 0}\left(e^{\frac{i}{\hbar}\vec k_{j}\vec x}-e^{\frac{i}{\hbar}\vec k_{j}\vec y}\right)\\\label{CCR1X}&\times\left(
e^{-\frac{i}{\hbar}\gamma_{j}t}\frac{c_{j}{\vec k_{j}}^{2}}{4m\gamma_{j}^{2}}\left((1+\tilde v_{j})\frac{\gamma_{j}}{\omega}+\tilde v_{j}\right)\hat a_{j}^{}
+e^{\frac{i}{\hbar}\gamma_{j}t}\frac{d_{j}{\vec k_{j}}^{2}}{4m\gamma_{j}^{2}}\left((1+\tilde v_{j})\frac{\gamma_{j}}{\omega}-\tilde v_{j}\right)\hat a_{-j}^{\dagger}
\right)
\end{align}
and
\begin{align}\nonumber
&[\hat\phi_{\times}(t,\vec x),\hat\varphi^{\dagger}(t,\vec y)]+[\hat\varphi(t,\vec x),\hat\phi_{\times}^{\dagger}(t,\vec y)]\\\nonumber &=\frac{\omega}{L^{3}}\sum\limits_{j\neq 0}\frac{c_{j}\left(\left(\frac{{\vec k_{j}}^{2}}{2m}\right)^{2}+\omega\tilde v_{j}\frac{{\vec k_{j}}^{2}}{2m}+\gamma_{j}^{2}+\frac{{\vec k_{j}}^{2}}{2m}\gamma_{j}(\tilde v_{j}-1)\right)}{2\omega\gamma_{j}^{2}}\left(e^{\frac{i}{\hbar}\vec k_{j}\vec x}
e^{-\frac{i}{\hbar}\gamma_{j}t}\hat a_{j}^{}+e^{-\frac{i}{\hbar}\vec k_{j}\vec y}e^{\frac{i}{\hbar}\gamma_{j}t}\hat a_{j}^{\dagger}\right)\\\label{CCR2X}
&-\frac{\omega}{L^{3}}\sum\limits_{j\neq 0}\frac{d_{j}\left(\left(\frac{{\vec k_{j}}^{2}}{2m}\right)^{2}+\omega\tilde v_{j}\frac{{\vec k_{j}}^{2}}{2m}+\gamma_{j}^{2}-\frac{{\vec k_{j}}^{2}}{2m}\gamma_{j}(\tilde v_{j}-1)\right)}{2\omega\gamma_{j}^{2}}\left(e^{\frac{i}{\hbar}\vec k_{j}\vec x}
e^{\frac{i}{\hbar}\gamma_{j}t}\hat a_{-j}^{\dagger}+e^{-\frac{i}{\hbar}\vec k_{j}\vec y}e^{-\frac{i}{\hbar}\gamma_{j}t}\hat a_{-j}^{}\right).
\end{align}
Although these expressions are not equal to zero, they will be compensated by contributions emerging from the oscillation modes.

\subsection{Operators $\hat\phi_{\textrm{o}}^{}(t,\vec x)$ for the oscillation modes}
\subsubsection{Separation of the equations of motion for the oscillation modes}
Using \eqref{Vexp}, the r.h.s. of Eq.~\eqref{eqquant3} can be brought to the explicit form
\begin{align}\nonumber
&\frac{\omega\sqrt{L^{3}}}{g}\int d^{3}yV(|\vec x-\vec y|)\left[\left(\hat\varphi_{\textrm{o}}^{}(t,\vec y)+\hat\varphi_{\textrm{o}}^{\dagger}(t,\vec y)\right)\hat\varphi_{\textrm{o}}^{}(t,\vec x)+\hat\varphi_{\textrm{o}}^{\dagger}(t,\vec y)\hat\varphi_{\textrm{o}}^{}(t,\vec y)\right]\\\nonumber&=\frac{\omega}{\sqrt{L^{3}}}\sum\limits_{j\neq 0}\left(\left((1+\tilde v_{j})d_{j}^{2}-\tilde v_{j}c_{j}d_{j}\right)\hat a_{j}^{}\hat a_{j}^{\dagger}+\left((1+\tilde v_{j})c_{j}^{2}-\tilde v_{j}c_{j}d_{j}\right)\hat a_{j}^{\dagger}\hat a_{j}^{}\right.\\\nonumber&\left.
+\,e^{-\frac{2i}{\hbar}\gamma_{j}t}\left(\tilde v_{j}c_{j}^{2}-(1+\tilde v_{j})c_{j}d_{j}\right)\hat a_{j}^{}\hat a_{-j}^{}
+e^{\frac{2i}{\hbar}\gamma_{j}t}\left(\tilde v_{j}d_{j}^{2}-(1+\tilde v_{j})c_{j}d_{j}\right)\hat a_{j}^{\dagger}\hat a_{-j}^{\dagger}\right)
\\\nonumber
&+\frac{\omega}{2\sqrt{L^{3}}}\underset{j\neq -l}{\sum\limits_{j\neq 0}\sum\limits_{l\neq 0}}\left(q_{j,l}^{(1)}e^{\frac{i}{\hbar}(\vec k_{j}+\vec k_{l})\vec x}e^{-\frac{i}{\hbar}(\gamma_{j}+\gamma_{l})t}\hat a_{j}^{}\hat a_{l}^{}+q_{j,l}^{(2)}e^{-\frac{i}{\hbar}(\vec k_{j}+\vec k_{l})\vec x}e^{\frac{i}{\hbar}(\gamma_{j}+\gamma_{l})t}\hat a_{j}^{\dagger}\hat a_{l}^{\dagger}\right)\\\label{oscrhs}
&+\frac{\omega}{2\sqrt{L^{3}}}\underset{j\neq l}{\sum\limits_{j\neq 0}\sum\limits_{l\neq 0}}\left(
q_{j,l}^{(3)}e^{\frac{i}{\hbar}(\vec k_{j}-\vec k_{l})\vec x}e^{-\frac{i}{\hbar}(\gamma_{j}-\gamma_{l})t}\hat a_{j}^{}\hat a_{l}^{\dagger}+q_{j,l}^{(4)}e^{-\frac{i}{\hbar}(\vec k_{j}-\vec k_{l})\vec x}e^{\frac{i}{\hbar}(\gamma_{j}-\gamma_{l})t}\hat a_{j}^{\dagger}\hat a_{l}^{}\right),
\end{align}
where
\begin{align}\label{qdef1}
&q_{j,l}^{(1)}=(\tilde v_{j}+\tilde v_{l})c_{j}c_{l}-(\tilde v_{l}+\tilde v_{j+l})c_{j}d_{l}-(\tilde v_{j}+\tilde v_{j+l})d_{j}c_{l},\\\label{qdef2}
&q_{j,l}^{(2)}=(\tilde v_{j}+\tilde v_{l})d_{j}d_{l}-(\tilde v_{j}+\tilde v_{j+l})c_{j}d_{l}-(\tilde v_{l}+\tilde v_{j+l})d_{j}c_{l},\\\label{qdef3}
&q_{j,l}^{(3)}=(\tilde v_{j}+\tilde v_{j-l})d_{j}d_{l}+(\tilde v_{l}+\tilde v_{j-l})c_{j}c_{l}-(\tilde v_{j}+\tilde v_{l})c_{j}d_{l},\\\label{qdef4}
&q_{j,l}^{(4)}=(\tilde v_{l}+\tilde v_{j-l})d_{j}d_{l}+(\tilde v_{j}+\tilde v_{j-l})c_{j}c_{l}-(\tilde v_{j}+\tilde v_{l})d_{j}c_{l}.
\end{align}
In order to get \eqref{oscrhs}, the first representation in \eqref{linsolquant} was used. Exactly as it was done in \cite{Smolyakov:2021vlo}, in representation \eqref{oscrhs} the terms that do not depend on $\vec x$ (the first sum), the terms that depend on $e^{\pm\frac{i}{\hbar}(\vec k_{j}+\vec k_{l})\vec x}$ (the first double sum), and the terms that depend on $e^{\pm\frac{i}{\hbar}(\vec k_{j}-\vec k_{l})\vec x}$ (the second double sum) are separated. The factor $\frac{1}{2}$ in the terms with double sums in \eqref{oscrhs} is due to symmetrization. Note that
\begin{equation}
q_{l,j}^{(1)}=q_{j,l}^{(1)},\qquad q_{l,j}^{(2)}=q_{j,l}^{(2)},\qquad q_{l,j}^{(3)}=q_{j,l}^{(4)}.
\end{equation}
Contrary to the case of \cite{Smolyakov:2021vlo}, here $q_{l,j}^{({\scriptscriptstyle\#})}\neq q_{j,-l}^{({\scriptscriptstyle\#})}$ because of the factors $\tilde v_{j\pm l}$.

It is useful to represent the operator $\hat\phi_{\textrm{o}}(t,\vec x)$ as it was done in \cite{Smolyakov:2021vlo}:
\begin{equation}\label{oscrepresent1}
\hat\phi_{\textrm{o}}(t,\vec x)=\hat\phi_{\textrm{t}}(t)+\frac{1}{2}\underset{\substack{j\neq 0, l\neq 0\\j\neq-l}}{\sum\sum}\hat\phi_{j,l}^{\scriptscriptstyle+}(t,\vec x)+\frac{1}{2}\underset{\substack{j\neq 0, l\neq 0\\j\neq l}}{\sum\sum}\hat\phi_{j,l}^{\scriptscriptstyle-}(t,\vec x),
\end{equation}
where the operators $\hat\phi_{\textrm{t}}(t)$, $\hat\phi_{j,l}^{\scriptscriptstyle+}(t,\vec x)$ and $\hat\phi_{j,l}^{\scriptscriptstyle-}(t,\vec x)$ in \eqref{oscrepresent1} satisfy the equations
\begingroup
\allowdisplaybreaks
\begin{align}\nonumber
&i\hbar\dot{\hat\phi}_{\textrm{t}}^{}-\omega\left(\hat\phi_{\textrm{t}}^{}+\hat\phi_{\textrm{t}}^{\dagger}\right)=\frac{\omega}{\sqrt{L^{3}}}\sum\limits_{j\neq 0}\left(\left((1+\tilde v_{j})d_{j}^{2}-\tilde v_{j}c_{j}d_{j}\right)\hat a_{j}^{}\hat a_{j}^{\dagger}+\left((1+\tilde v_{j})c_{j}^{2}-\tilde v_{j}c_{j}d_{j}\right)\hat a_{j}^{\dagger}\hat a_{j}^{}\right.\\\label{eqonlyt}&\left.
+\,e^{-\frac{2i}{\hbar}\gamma_{j}t}\left(\tilde v_{j}c_{j}^{2}-(1+\tilde v_{j})c_{j}d_{j}\right)\hat a_{j}^{}\hat a_{-j}^{}
+e^{\frac{2i}{\hbar}\gamma_{j}t}\left(\tilde v_{j}d_{j}^{2}-(1+\tilde v_{j})c_{j}d_{j}\right)\hat a_{j}^{\dagger}\hat a_{-j}^{\dagger}\right),\\\nonumber
&i\hbar\dot{\hat\phi}_{j,l}^{\scriptscriptstyle+}(t,\vec x)+\frac{\hbar^{2}}{2m}\Delta\hat\phi_{j,l}^{\scriptscriptstyle+}(t,\vec x)-\frac{\omega}{g}\int d^{3}yV(|\vec x-\vec y|)\left(\hat\phi_{j,l}^{\scriptscriptstyle+}(t,\vec y)
+\hat\phi_{j,l}^{\scriptscriptstyle+\dagger}(t,\vec y)\right)\\\label{eqosc+}
&=\frac{\omega}{\sqrt{L^{3}}}\left(q_{j,l}^{(1)}e^{\frac{i}{\hbar}(\vec k_{j}+\vec k_{l})\vec x}e^{-\frac{i}{\hbar}(\gamma_{j}+\gamma_{l})t}\hat a_{j}^{}\hat a_{l}^{}+q_{j,l}^{(2)}e^{-\frac{i}{\hbar}(\vec k_{j}+\vec k_{l})\vec x}e^{\frac{i}{\hbar}(\gamma_{j}+\gamma_{l})t}\hat a_{j}^{\dagger}\hat a_{l}^{\dagger}\right),\\\nonumber
&i\hbar\dot{\hat\phi}_{j,l}^{\scriptscriptstyle-}(t,\vec x)+\frac{\hbar^{2}}{2m}\Delta\hat\phi_{j,l}^{\scriptscriptstyle-}(t,\vec x)-\frac{\omega}{g}\int d^{3}yV(|\vec x-\vec y|)\left(\hat\phi_{j,l}^{\scriptscriptstyle-}(t,\vec y)
+\hat\phi_{j,l}^{\scriptscriptstyle-\dagger}(t,\vec y)\right)\\\label{eqosc-}
&=\frac{\omega}{\sqrt{L^{3}}}\left(q_{j,l}^{(3)}e^{\frac{i}{\hbar}(\vec k_{j}-\vec k_{l})\vec x}e^{-\frac{i}{\hbar}(\gamma_{j}-\gamma_{l})t}\hat a_{j}^{}\hat a_{l}^{\dagger}+q_{j,l}^{(4)}e^{-\frac{i}{\hbar}(\vec k_{j}-\vec k_{l})\vec x}e^{\frac{i}{\hbar}(\gamma_{j}-\gamma_{l})t}\hat a_{j}^{\dagger}\hat a_{l}^{}\right).
\end{align}
\endgroup
Now we are ready to solve Eqs.~\eqref{eqonlyt}--\eqref{eqosc-}.

\subsubsection{Operator $\hat\phi_{\textrm{t}}(t)$}\label{sectCompensation}
Explicit solution of Eq.~\eqref{eqonlyt} looks as
\begin{align}\nonumber
\hat\phi_{\textrm{t}}(t)=\frac{\omega}{\sqrt{L^{3}}}\sum\limits_{l\neq 0}&\left(-\frac{\frac{1+\tilde v_{j}}{2}\left(\frac{{\vec k_{j}}^{2}}{2m}+\gamma_{j}\right)+\frac{\omega\tilde v_{j}}{2}}{2\omega\gamma_{j}}
\hat a_{j}^{\dagger}\hat a_{j}^{}-\frac{\frac{1+\tilde v_{j}}{2}\left(\frac{{\vec k_{j}}^{2}}{2m}-\gamma_{j}\right)+\frac{\omega\tilde v_{j}}{2}}{2\omega\gamma_{j}}\hat a_{j}^{}\hat a_{j}^{\dagger}\right.\\\label{phitosc}&\left.
+\frac{\tilde v_{j}\left(\gamma_{j}+\frac{{\vec k_{j}}^{2}}{2m}\right)}{4\gamma_{j}^{2}}e^{-\frac{2i}{\hbar}\gamma_{j}t}\hat a_{j}^{}\hat a_{-j}^{}
+\frac{\tilde v_{j}\left(\gamma_{j}-\frac{{\vec k_{j}}^{2}}{2m}\right)}{4\gamma_{j}^{2}}e^{\frac{2i}{\hbar}\gamma_{j}t}\hat a_{j}^{\dagger}\hat a_{-j}^{\dagger}\right).
\end{align}
For this solution, commutation relations \eqref{CCR1} and \eqref{CCR2} take the form
\begin{align}\nonumber
&[\hat\phi_{\textrm{t}}(t),\hat\varphi(t,\vec y)]+[\hat\varphi(t,\vec x),\hat\phi_{\textrm{t}}(t)]
=\frac{\omega}{L^{3}}\sum\limits_{j\neq 0}\left(e^{\frac{i}{\hbar}\vec k_{j}\vec y}-e^{\frac{i}{\hbar}\vec k_{j}\vec x}\right)\\\label{CCR1ot}&\times\left(
e^{-\frac{i}{\hbar}\gamma_{j}t}\frac{c_{j}{\vec k_{j}}^{2}}{4m\gamma_{j}^{2}}\left((1+\tilde v_{j})\frac{\gamma_{j}}{\omega}+\tilde v_{j}\right)\hat a_{j}^{}
+e^{\frac{i}{\hbar}\gamma_{j}t}\frac{d_{j}{\vec k_{j}}^{2}}{4m\gamma_{j}^{2}}\left((1+\tilde v_{j})\frac{\gamma_{j}}{\omega}-\tilde v_{j}\right)\hat a_{-j}^{\dagger}
\right)
\end{align}
and
\begin{align}\nonumber
&[\hat\phi_{\textrm{t}}(t),\hat\varphi^{\dagger}(t,\vec y)]+[\hat\varphi(t,\vec x),\hat\phi_{\textrm{t}}^{\dagger}(t)]\\\nonumber
&=\frac{\omega}{L^{3}}\sum\limits_{j\neq 0}\frac{d_{j}\left(\left(\frac{{\vec k_{j}}^{2}}{2m}\right)^{2}+\omega\tilde v_{j}\frac{{\vec k_{j}}^{2}}{2m}+\gamma_{j}^{2}-\frac{{\vec k_{j}}^{2}}{2m}\gamma_{j}(\tilde v_{j}-1)\right)}{2\omega\gamma_{j}^{2}}\left(e^{\frac{i}{\hbar}\vec k_{j}\vec x}
e^{\frac{i}{\hbar}\gamma_{j}t}\hat a_{-j}^{\dagger}+e^{-\frac{i}{\hbar}\vec k_{j}\vec y}e^{-\frac{i}{\hbar}\gamma_{j}t}\hat a_{-j}^{}\right)\\\label{CCR2ot}
&-\frac{\omega}{L^{3}}\sum\limits_{j\neq 0}\frac{c_{j}\left(\left(\frac{{\vec k_{j}}^{2}}{2m}\right)^{2}+\omega\tilde v_{j}\frac{{\vec k_{j}}^{2}}{2m}+\gamma_{j}^{2}+\frac{{\vec k_{j}}^{2}}{2m}\gamma_{j}(\tilde v_{j}-1)\right)}{2\omega\gamma_{j}^{2}}\left(e^{\frac{i}{\hbar}\vec k_{j}\vec x}
e^{-\frac{i}{\hbar}\gamma_{j}t}\hat a_{j}^{}+e^{-\frac{i}{\hbar}\vec k_{j}\vec y}e^{\frac{i}{\hbar}\gamma_{j}t}\hat a_{j}^{\dagger}\right).
\end{align}
One can see that \eqref{CCR1ot} fully compensates \eqref{CCR1X}, and \eqref{CCR2ot} fully compensates \eqref{CCR2X}, exactly as it happens in the simpler case of \eqref{interactpotent} \cite{Smolyakov:2021vlo}.

\subsubsection{Operators $\hat\phi_{j,l}^{\scriptscriptstyle\pm}(t,\vec x)$}
It is clear that for the most $j$ and $l$ the relations
\begin{align}\label{notransform1}
&(\gamma_{j}+\gamma_{l})^{2}-\frac{(\vec k_{j}+\vec k_{l})^{2}}{2m}\left(\frac{(\vec k_{j}+\vec k_{l})^{2}}{2m}+2\omega\tilde v_{j+l}\right)\neq 0,\\\label{notransform2}
&(\gamma_{j}-\gamma_{l})^{2}-\frac{(\vec k_{j}-\vec k_{l})^{2}}{2m}\left(\frac{(\vec k_{j}-\vec k_{l})^{2}}{2m}+2\omega\tilde v_{j-l}\right)\neq 0
\end{align}
are fulfilled. In this case, one can take solutions of Eqs.~\eqref{eqosc+},~\eqref{eqosc-} in the following form \cite{Smolyakov:2021vlo}:
\begin{align}\label{phijldef+}
&\hat\phi_{j,l}^{\scriptscriptstyle+}(t,\vec x)=\frac{\omega}{\sqrt{L^{3}}}\left(M_{j,l}^{\star\scriptscriptstyle+}e^{\frac{i}{\hbar}(\vec k_{j}+\vec k_{l})\vec x}e^{-\frac{i}{\hbar}(\gamma_{j}+\gamma_{l})t}\hat a_{j}^{}\hat a_{l}^{}+N_{j,l}^{\star\scriptscriptstyle+}e^{-\frac{i}{\hbar}(\vec k_{j}+\vec k_{l})\vec x}e^{\frac{i}{\hbar}(\gamma_{j}+\gamma_{l})t}\hat a_{j}^{\dagger}\hat a_{l}^{\dagger}\right),\\\label{phijldef-}
&\hat\phi_{j,l}^{\scriptscriptstyle-}(t,\vec x)=\frac{\omega}{\sqrt{L^{3}}}\left(M_{j,l}^{\star\scriptscriptstyle-}e^{\frac{i}{\hbar}(\vec k_{j}-\vec k_{l})\vec x}e^{-\frac{i}{\hbar}(\gamma_{j}-\gamma_{l})t}\hat a_{j}^{}\hat a_{l}^{\dagger}+N_{j,l}^{\star\scriptscriptstyle-}e^{-\frac{i}{\hbar}(\vec k_{j}-\vec k_{l})\vec x}e^{\frac{i}{\hbar}(\gamma_{j}-\gamma_{l})t}\hat a_{j}^{\dagger}\hat a_{l}^{}\right).
\end{align}
Substituting \eqref{phijldef+} into Eq.~\eqref{eqosc+}, we get the algebraic system of equations
\begin{align}\label{tildeNeq+}
&M_{j,l}^{\star\scriptscriptstyle+}\left(\frac{(\vec k_{j}+\vec k_{l})^{2}}{2m}+\omega\tilde v_{j+l}-(\gamma_{j}+\gamma_{l})\right)+\omega\tilde v_{j+l} \left(N_{j,l}^{\star\scriptscriptstyle+}\right)^{*}=-q_{j,l}^{(1)},\\
&M_{j,l}^{\star\scriptscriptstyle+}\omega\tilde v_{j+l}+\left(\frac{(\vec k_{j}+\vec k_{l})^{2}}{2m}+\omega\tilde v_{j+l}+(\gamma_{j}+\gamma_{l})\right) \left(N_{j,l}^{\star\scriptscriptstyle+}\right)^{*}=-q_{j,l}^{(2)}
\end{align}
for the coefficients in \eqref{phijldef+}, leading to
\begin{align}\label{coefftildeM+}
&M_{j,l}^{\star\scriptscriptstyle+}=\frac{\left(\frac{(\vec k_{j}+\vec k_{l})^{2}}{2m}+(\gamma_{j}+\gamma_{l})\right)q_{j,l}^{(1)}+\omega\tilde v_{j+l}\left( q_{j,l}^{(1)}-q_{j,l}^{(2)}\right)}
{(\gamma_{j}+\gamma_{l})^{2}+(\omega\tilde v_{j+l})^{2}-\left(\frac{(\vec k_{j}+\vec k_{l})^{2}}{2m}+\omega\tilde v_{j+l}\right)^{2}},\\\label{coefftildeN+}
&N_{j,l}^{\star\scriptscriptstyle+}=\frac{\left(\frac{(\vec k_{j}+\vec k_{l})^{2}}{2m}-(\gamma_{j}+\gamma_{l})\right)q_{j,l}^{(2)}-\omega\tilde v_{j+l}\left( q_{j,l}^{(1)}-q_{j,l}^{(2)}\right)}
{(\gamma_{j}+\gamma_{l})^{2}+(\omega\tilde v_{j+l})^{2}-\left(\frac{(\vec k_{j}+\vec k_{l})^{2}}{2m}+\omega\tilde v_{j+l}\right)^{2}}.
\end{align}
Substituting \eqref{phijldef-} into Eq.~\eqref{eqosc-}, we get the algebraic system of equations
\begin{align}\label{tildeNeq-}
&M_{j,l}^{\star\scriptscriptstyle-}\left(\frac{(\vec k_{j}-\vec k_{l})^{2}}{2m}+\omega\tilde v_{j-l}-(\gamma_{j}-\gamma_{l})\right)+\omega\tilde v_{j-l}\left(N_{j,l}^{\star\scriptscriptstyle-}\right)^{*}=-q_{j,l}^{(3)},\\
&M_{j,l}^{\star\scriptscriptstyle-}\omega\tilde v_{j-l}+\left(\frac{(\vec k_{j}-\vec k_{l})^{2}}{2m}+\omega\tilde v_{j-l}+(\gamma_{j}-\gamma_{l})\right) \left(N_{j,l}^{\star\scriptscriptstyle-}\right)^{*}=-q_{j,l}^{(4)}
\end{align}
for the coefficients in \eqref{phijldef-}, leading to
\begin{align}\label{coefftildeM-}
&M_{j,l}^{\star\scriptscriptstyle-}=\frac{\left(\frac{(\vec k_{j}-\vec k_{l})^{2}}{2m}+(\gamma_{j}-\gamma_{l})\right)q_{j,l}^{(3)}+\omega\tilde v_{j-l}\left( q_{j,l}^{(3)}-q_{j,l}^{(4)}\right)}
{(\gamma_{j}-\gamma_{l})^{2}+(\omega\tilde v_{j-l})^{2}-\left(\frac{(\vec k_{j}-\vec k_{l})^{2}}{2m}+\omega\tilde v_{j-l}\right)^{2}},\\\label{coefftildeN-}
&N_{j,l}^{\star\scriptscriptstyle-}=\frac{\left(\frac{(\vec k_{j}-\vec k_{l})^{2}}{2m}-(\gamma_{j}-\gamma_{l})\right)q_{j,l}^{(4)}-\omega\tilde v_{j-l}\left( q_{j,l}^{(3)}-q_{j,l}^{(4)}\right)}
{(\gamma_{j}-\gamma_{l})^{2}+(\omega\tilde v_{j-l})^{2}-\left(\frac{(\vec k_{j}-\vec k_{l})^{2}}{2m}+\omega\tilde v_{j-l}\right)^{2}}.
\end{align}
Coefficients \eqref{coefftildeM+}, \eqref{coefftildeN+} and \eqref{coefftildeM-}, \eqref{coefftildeN-} are such that
\begin{equation}\label{starMNsymmetry}
M_{l,j}^{\star\scriptscriptstyle+}=M_{j,l}^{\star\scriptscriptstyle+},\quad N_{l,j}^{\star\scriptscriptstyle+}=N_{j,l}^{\star\scriptscriptstyle+},\quad  N_{l,j}^{\star\scriptscriptstyle-}=M_{j,l}^{\star\scriptscriptstyle-}.
\end{equation}

However, in principle there may exist some $j$ and $l$ such that for $k=j+l$ (i.e., $\vec k_{k}=\vec k_{j}+\vec k_{l}$) the frequency $\gamma_{k}=\gamma_{j+l}=\gamma_{j}+\gamma_{l}$ (analogously, for $\gamma_{j-l}=|\gamma_{j}-\gamma_{l}|$). In other words, it is possible that for such $j$ and $l$ the equality
\begin{equation}\label{transform1}
(\gamma_{j}+\gamma_{l})^{2}-\frac{(\vec k_{j}+\vec k_{l})^{2}}{2m}\left(\frac{(\vec k_{j}+\vec k_{l})^{2}}{2m}+2\omega\tilde v_{j+l}\right)=0
\end{equation}
holds or the equality
\begin{equation}\label{transform2}
(\gamma_{j}-\gamma_{l})^{2}-\frac{(\vec k_{j}-\vec k_{l})^{2}}{2m}\left(\frac{(\vec k_{j}-\vec k_{l})^{2}}{2m}+2\omega\tilde v_{j-l}\right)=0
\end{equation}
holds (see an example in the limit $L\to\infty$ in Appendix~A of \cite{Smolyakov:2021vlo}). For such $j$ and $l$, one should take solutions of Eqs.~\eqref{eqosc+},~\eqref{eqosc-} in a different form \cite{Smolyakov:2021vlo}:
\begin{align}\nonumber
\hat\phi_{j,l}^{\scriptscriptstyle+}(t,\vec x)=\frac{\omega}{L^{\frac{3}{2}}}&\Bigl(\left(M_{j,l}^{\scriptscriptstyle+}+tL_{j,l}^{\scriptscriptstyle+}\right)e^{\frac{i}{\hbar}(\vec k_{j}+\vec k_{l})\vec x}e^{-\frac{i}{\hbar}(\gamma_{j}+\gamma_{l})t}\hat a_{j}^{}\hat a_{l}^{}\\\label{phijldef+1}&+\left(N_{j,l}^{\scriptscriptstyle+}+tJ_{j,l}^{\scriptscriptstyle+}\right)e^{-\frac{i}{\hbar}(\vec k_{j}+\vec k_{l})\vec x}e^{\frac{i}{\hbar}(\gamma_{j}+\gamma_{l})t}\hat a_{j}^{\dagger}\hat a_{l}^{\dagger}\Bigr),\\\nonumber
\hat\phi_{j,l}^{\scriptscriptstyle-}(t,\vec x)=\frac{\omega}{L^{\frac{3}{2}}}&\Bigl(\left(M_{j,l}^{\scriptscriptstyle-}+tL_{j,l}^{\scriptscriptstyle-}\right)e^{\frac{i}{\hbar}(\vec k_{j}-\vec k_{l})\vec x}e^{-\frac{i}{\hbar}(\gamma_{j}-\gamma_{l})t}\hat a_{j}^{}\hat a_{l}^{\dagger}\\\label{phijldef-1}&+\left(N_{j,l}^{\scriptscriptstyle-}+tJ_{j,l}^{\scriptscriptstyle-}\right)e^{-\frac{i}{\hbar}(\vec k_{j}-\vec k_{l})\vec x}e^{\frac{i}{\hbar}(\gamma_{j}-\gamma_{l})t}\hat a_{j}^{\dagger}\hat a_{l}^{}\Bigr).
\end{align}
Substituting \eqref{phijldef+1} into Eq.~\eqref{eqosc+}, we get the algebraic system of equations
\begin{align}\label{LJeq+}
&L_{j,l}^{\scriptscriptstyle+}\left(\frac{(\vec k_{j}+\vec k_{l})^{2}}{2m}+\omega\tilde v_{j+l}-(\gamma_{j}+\gamma_{l})\right)+\omega\tilde v_{j+l} \left(J_{j,l}^{\scriptscriptstyle+}\right)^{*}=0,\\
&L_{j,l}^{\scriptscriptstyle+}\omega\tilde v_{j+l}+\left(\frac{(\vec k_{j}+\vec k_{l})^{2}}{2m}+\omega\tilde v_{j+l}+(\gamma_{j}+\gamma_{l})\right)\left(J_{j,l}^{\scriptscriptstyle+}\right)^{*}=0,\\
&M_{j,l}^{\scriptscriptstyle+}\left(\frac{(\vec k_{j}+\vec k_{l})^{2}}{2m}+\omega\tilde v_{j+l}-(\gamma_{j}+\gamma_{l})\right)+\omega\tilde v_{j+l}\left(N_{j,l}^{\scriptscriptstyle+}\right)^{*}-i\hbar L_{j,l}^{\scriptscriptstyle+}=-q_{j,l}^{(1)},\\
&M_{j,l}^{\scriptscriptstyle+}\omega\tilde v_{j+l}+\left(\frac{(\vec k_{j}+\vec k_{l})^{2}}{2m}+\omega\tilde v_{j+l}+(\gamma_{j}+\gamma_{l})\right)\left(N_{j,l}^{\scriptscriptstyle+}\right)^{*}+i\hbar\left(J_{j,l}^{\scriptscriptstyle+}\right)^{*}=-q_{j,l}^{(2)}
\end{align}
for the coefficients in \eqref{phijldef+1}. It is useful to choose the following solution of this system of equations:
\begin{align}\label{L+def}
&L_{j,l}^{\scriptscriptstyle+}=\frac{-i}{2\hbar(\gamma_{j}+\gamma_{l})}\left(\left(\frac{(\vec k_{j}+\vec k_{l})^{2}}{2m}+(\gamma_{j}+\gamma_{l})\right)q_{j,l}^{(1)}+\omega\tilde v_{j+l}\left(q_{j,l}^{(1)}-q_{j,l}^{(2)}\right)\right),\\\label{J+def}
&J_{j,l}^{\scriptscriptstyle+}=\frac{i}{2\hbar(\gamma_{j}+\gamma_{l})}\left(\left(\frac{(\vec k_{j}+\vec k_{l})^{2}}{2m}-(\gamma_{j}+\gamma_{l})\right)q_{j,l}^{(2)}-\omega\tilde v_{j+l}\left(q_{j,l}^{(1)}-q_{j,l}^{(2)}\right)\right),\\\label{M+def}
&M_{j,l}^{\scriptscriptstyle+}=K_{j,l}^{\scriptscriptstyle+},\\
\label{N+def}
&N_{j,l}^{\scriptscriptstyle+}=0,
\end{align}
where
\begin{equation}\label{Kab+}
K_{j,l}^{\scriptscriptstyle+}=\frac{-1}{2\omega\tilde v_{j+l}(\gamma_{j}+\gamma_{l})}\left(\left(\frac{(\vec k_{j}+\vec k_{l})^{2}}{2m}+(\gamma_{j}+\gamma_{l})\right)q_{j,l}^{(2)}-\omega\tilde v_{j+l}\left(q_{j,l}^{(1)}-q_{j,l}^{(2)}\right)\right)
\end{equation}
such that $K_{j,l}^{\scriptscriptstyle+}=K_{l,j}^{\scriptscriptstyle+}$.

Substituting \eqref{phijldef-1} into Eq.~\eqref{eqosc-}, we get the algebraic system of equations
\begin{align}\label{LJeq-}
&L_{j,l}^{\scriptscriptstyle-}\left(\frac{(\vec k_{j}-\vec k_{l})^{2}}{2m}+\omega\tilde v_{j-l}-(\gamma_{j}-\gamma_{l})\right)+\omega\tilde v_{j-l}\left(J_{j,l}^{\scriptscriptstyle-}\right)^{*}=0,\\
&L_{j,l}^{\scriptscriptstyle-}\omega\tilde v_{j-l}+\left(\frac{(\vec k_{j}-\vec k_{l})^{2}}{2m}+\omega\tilde v_{j-l}+(\gamma_{j}-\gamma_{l})\right)\left(J_{j,l}^{\scriptscriptstyle-}\right)^{*}=0,\\
&M_{j,l}^{\scriptscriptstyle-}\left(\frac{(\vec k_{j}-\vec k_{l})^{2}}{2m}+\omega\tilde v_{j-l}-(\gamma_{j}-\gamma_{l})\right)+\omega\tilde v_{j-l}\left(N_{j,l}^{\scriptscriptstyle-}\right)^{*}-i\hbar L_{j,l}^{\scriptscriptstyle-}=-q_{j,l}^{(3)},\\
&M_{j,l}^{\scriptscriptstyle-}\omega\tilde v_{j-l}+\left(\frac{(\vec k_{j}-\vec k_{l})^{2}}{2m}+\omega\tilde v_{j-l}+(\gamma_{j}-\gamma_{l})\right)\left(N_{j,l}^{\scriptscriptstyle-}\right)^{*}+i\hbar\left(J_{j,l}^{\scriptscriptstyle-}\right)^{*}=-q_{j,l}^{(4)}
\end{align}
for the coefficients in \eqref{phijldef-1}. It is useful to choose the following solution of this system of equations:
\begin{align}\label{L-def}
&L_{j,l}^{\scriptscriptstyle-}=\frac{-i}{2\hbar(\gamma_{j}-\gamma_{l})}\left(\left(\frac{(\vec k_{j}-\vec k_{l})^{2}}{2m}+(\gamma_{j}-\gamma_{l})\right)q_{j,l}^{(3)}+\omega\tilde v_{j-l}\left(q_{j,l}^{(3)}-q_{j,l}^{(4)}\right)\right),\\\label{J-def}
&J_{j,l}^{\scriptscriptstyle-}=\frac{i}{2\hbar(\gamma_{j}-\gamma_{l})}\left(\left(\frac{(\vec k_{j}-\vec k_{l})^{2}}{2m}-(\gamma_{j}-\gamma_{l})\right)q_{j,l}^{(4)}-\omega\tilde v_{j-l}\left(q_{j,l}^{(3)}-q_{j,l}^{(4)}\right)\right),\\\label{M-def}
&M_{j,l}^{\scriptscriptstyle-}=K_{j,l}^{\scriptscriptstyle-}+c_{j-l}Q_{j,l}^{\scriptscriptstyle-},\\
\label{N-def}
&N_{j,l}^{\scriptscriptstyle-}=-d_{j-l}Q_{j,l}^{\scriptscriptstyle-},
\end{align}
where
\begin{equation}\label{Kab-}
K_{j,l}^{\scriptscriptstyle-}=\frac{-1}{2\omega\tilde v_{j-l}(\gamma_{j}-\gamma_{l})}\left(\left(\frac{(\vec k_{j}-\vec k_{l})^{2}}{2m}+(\gamma_{j}-\gamma_{l})\right)q_{j,l}^{(4)}-\omega\tilde v_{j-l}\left(q_{j,l}^{(3)}-q_{j,l}^{(4)}\right)\right)
\end{equation}
and the coefficients $Q_{j,l}^{\scriptscriptstyle-}$ will be specified later.

Now it is very useful to represent the double sums in \eqref{oscrepresent1} as it was done in \cite{Smolyakov:2021vlo}:
\begin{align}\nonumber
&\frac{1}{2}\underset{\substack{j\neq 0, l\neq 0\\j\neq-l}}{\sum\sum}\hat\phi_{j,l}^{\scriptscriptstyle+}(t,\vec x)+\frac{1}{2}\underset{\substack{j\neq 0, l\neq 0\\j\neq l}}{\sum\sum}\hat\phi_{j,l}^{\scriptscriptstyle-}(t,\vec x)\\\label{oscrepresent2}&=\frac{1}{2}\underset{\substack{j\neq 0, l\neq 0\\j\neq-l}}{\sum\sum}\hat{\tilde\phi}_{j,l}^{\scriptscriptstyle+}(t,\vec x)+
\frac{1}{2}\underset{\substack{j\neq 0, l\neq 0\\j\neq l}}{\sum\sum}\hat{\tilde\phi}_{j,l}^{\scriptscriptstyle-}(t,\vec x)+
\frac{1}{2}\underset{\substack{[j,l]}}{\sum\sum}\hat\rho_{j,l}^{\scriptscriptstyle+}(t,\vec x)+\frac{1}{2}\underset{\substack{(j,l)}}{\sum\sum}\hat\rho_{j,l}^{\scriptscriptstyle-}(t,\vec x).
\end{align}
Here the operators $\hat{\tilde\phi}_{j,l}^{\scriptscriptstyle\pm}(t,\vec x)$ have the form \cite{Smolyakov:2021vlo}
\begin{align}\label{phitildehat1}
&\hat{\tilde\phi}_{j,l}^{\scriptscriptstyle+}(t,\vec x)=\frac{\omega}{\sqrt{L^{3}}}\left(\tilde M_{j,l}^{\scriptscriptstyle+}e^{\frac{i}{\hbar}(\vec k_{j}+\vec k_{l})\vec x}e^{-\frac{i}{\hbar}(\gamma_{j}+\gamma_{l})t}\hat a_{j}^{}\hat a_{l}^{}+\tilde N_{j,l}^{\scriptscriptstyle+}e^{-\frac{i}{\hbar}(\vec k_{j}+\vec k_{l})\vec x}e^{\frac{i}{\hbar}(\gamma_{j}+\gamma_{l})t}\hat a_{j}^{\dagger}\hat a_{l}^{\dagger}\right),\\\label{phitildehat2}
&\hat{\tilde\phi}_{j,l}^{\scriptscriptstyle-}(t,\vec x)=\frac{\omega}{\sqrt{L^{3}}}\left(\tilde M_{j,l}^{\scriptscriptstyle-}e^{\frac{i}{\hbar}(\vec k_{j}-\vec k_{l})\vec x}e^{-\frac{i}{\hbar}(\gamma_{j}-\gamma_{l})t}\hat a_{j}^{}\hat a_{l}^{\dagger}+\tilde N_{j,l}^{\scriptscriptstyle-}e^{-\frac{i}{\hbar}(\vec k_{j}-\vec k_{l})\vec x}e^{\frac{i}{\hbar}(\gamma_{j}-\gamma_{l})t}\hat a_{j}^{\dagger}\hat a_{l}^{}\right).
\end{align}
The coefficients $\tilde M_{j,l}^{\scriptscriptstyle\pm}$ and $\tilde N_{j,l}^{\scriptscriptstyle\pm}$ in the latter representations are the following:
if $j$ and $l$ are such that relation \eqref{notransform1} holds, then $\tilde M_{j,l}^{\scriptscriptstyle+}=M_{j,l}^{\star\scriptscriptstyle+}$ and $\tilde N_{j,l}^{\scriptscriptstyle+}=N_{j,l}^{\star\scriptscriptstyle+}$; if $j$ and $l$ are such that relation \eqref{notransform2} holds, then
$\tilde M_{j,l}^{\scriptscriptstyle-}=M_{j,l}^{\star\scriptscriptstyle-}$ and $\tilde N_{j,l}^{\scriptscriptstyle-}=N_{j,l}^{\star\scriptscriptstyle-}$;
if $j$ and $l$ are such that relation \eqref{transform1} holds, then
$\tilde M_{j,l}^{\scriptscriptstyle+}=M_{j,l}^{\scriptscriptstyle+}$ and $\tilde N_{j,l}^{\scriptscriptstyle+}=N_{j,l}^{\scriptscriptstyle+}$;
and if $j$ and $l$ are such that relation \eqref{transform2} holds, then
$\tilde M_{j,l}^{\scriptscriptstyle-}=M_{j,l}^{\scriptscriptstyle-}$ and $\tilde N_{j,l}^{\scriptscriptstyle-}=N_{j,l}^{\scriptscriptstyle-}$. The operators $\hat\rho_{j,l}^{\scriptscriptstyle\pm}(t,\vec x)$ have the form \cite{Smolyakov:2021vlo}
\begin{align}\label{rho+}
&\hat\rho_{j,l}^{\scriptscriptstyle+}(t,\vec x)=\frac{\omega t}{\sqrt{L^{3}}}\left(L_{j,l}^{\scriptscriptstyle+}e^{\frac{i}{\hbar}(\vec k_{j}+\vec k_{l})\vec x}e^{-\frac{i}{\hbar}(\gamma_{j}+\gamma_{l})t}\hat a_{j}^{}\hat a_{l}^{}+J_{j,l}^{\scriptscriptstyle+}e^{-\frac{i}{\hbar}(\vec k_{j}+\vec k_{l})\vec x}e^{\frac{i}{\hbar}(\gamma_{j}+\gamma_{l})t}\hat a_{j}^{\dagger}\hat a_{l}^{\dagger}\right),\\\label{rho-}
&\hat\rho_{j,l}^{\scriptscriptstyle-}(t,\vec x)=\frac{\omega t}{\sqrt{L^{3}}}\left(L_{j,l}^{\scriptscriptstyle-}e^{\frac{i}{\hbar}(\vec k_{j}-\vec k_{l})\vec x}e^{-\frac{i}{\hbar}(\gamma_{j}-\gamma_{l})t}\hat a_{j}^{}\hat a_{l}^{\dagger}+J_{j,l}^{\scriptscriptstyle-}e^{-\frac{i}{\hbar}(\vec k_{j}-\vec k_{l})\vec x}e^{\frac{i}{\hbar}(\gamma_{j}-\gamma_{l})t}\hat a_{j}^{\dagger}\hat a_{l}^{}\right)
\end{align}
(note the overall factor $t$ in \eqref{rho+} and \eqref{rho-}). The notation $[j,l]$ in \eqref{oscrepresent2} labels the set of $j$ and $l$ for which relation \eqref{transform1} holds, the notation $(j,l)$ in \eqref{oscrepresent2} labels the set of $j$ and $l$ for which relation \eqref{transform2} holds. Now let us turn to the commutation relations.

\subsubsection{Commutation relations for the operators $\hat{\tilde\phi}_{j,l}^{\scriptscriptstyle\pm}(t,\vec x)$}
Since in terms of $M_{j,l}^{\star\scriptscriptstyle\pm}$, $N_{j,l}^{\star\scriptscriptstyle\pm}$ and $M_{j,l}^{\scriptscriptstyle\pm}$, $N_{j,l}^{\scriptscriptstyle\pm}$ the operators $\hat{\tilde\phi}_{j,l}^{\scriptscriptstyle\pm}(t,\vec x)$ defined by \eqref{phitildehat1} and \eqref{phitildehat2} have exactly the same form as those in \cite{Smolyakov:2021vlo}, we can use the expressions for the commutators obtained in \cite{Smolyakov:2021vlo}. Namely, for the operator
\begin{equation}\label{tildephidef}
\hat{\tilde\phi}^{\scriptscriptstyle}(t,\vec x)=\frac{1}{2}\underset{\substack{j\neq 0, l\neq 0\\j\neq-l}}{\sum\sum}\hat{\tilde\phi}_{j,l}^{\scriptscriptstyle+}(t,\vec x)+\frac{1}{2}\underset{\substack{j\neq 0, l\neq 0\\j\neq l}}{\sum\sum}\hat{\tilde\phi}_{j,l}^{\scriptscriptstyle-}(t,\vec x),
\end{equation}
the first commutation relation looks like \cite{Smolyakov:2021vlo}
\begin{align}\nonumber
&[\hat{\tilde\phi}(t,\vec x),\hat\varphi(t,\vec y)]+[\hat\varphi(t,\vec x),\hat{\tilde\phi}(t,\vec y)]\\\label{commtildefinal}
&=\frac{\omega}{L^{3}}\underset{\substack{j\neq 0, l\neq 0\\j\neq-l}}{\sum\sum}\Bigl(\hat a_{j}^{}e^{-\frac{i}{\hbar}\gamma_{j}t}e^{\frac{i}{\hbar}(\vec k_{j}+\vec k_{l})\vec y}e^{-\frac{i}{\hbar}\vec k_{l}\vec x}Y^{(1)}_{j,l}+\hat a_{j}^{\dagger}e^{\frac{i}{\hbar}\gamma_{j}t}e^{-\frac{i}{\hbar}(\vec k_{j}+\vec k_{l})\vec y}e^{\frac{i}{\hbar}\vec k_{l}\vec x}Y^{(2)}_{j,l}\Bigr),
\end{align}
whereas the second commutation relation looks like \cite{Smolyakov:2021vlo}
\begin{align}\nonumber
&[\hat{\tilde\phi}(t,\vec x),\hat\varphi^{\dagger}(t,\vec y)]+[\hat\varphi(t,\vec x),\hat{\tilde\phi}^{\dagger}(t,\vec y)]\\\label{commdagger}&
=\frac{\omega}{L^{3}}\underset{\substack{j\neq 0, l\neq 0\\j\neq-l}}{\sum\sum}\Bigl(\hat a_{j}^{\dagger}e^{\frac{i}{\hbar}\gamma_{j}t}e^{-\frac{i}{\hbar}(\vec k_{j}+\vec k_{l})\vec y}e^{\frac{i}{\hbar}\vec k_{l}\vec x}
+\hat a_{j}^{}e^{-\frac{i}{\hbar}\gamma_{j}t}e^{\frac{i}{\hbar}(\vec k_{j}+\vec k_{l})\vec x}e^{-\frac{i}{\hbar}\vec k_{l}\vec y}\Bigr)Y^{(3)}_{j,l}.
\end{align}
The coefficients $Y^{(1)}_{j,l}$, $Y^{(2)}_{j,l}$, $Y^{(3)}_{j,l}$ in \eqref{commtildefinal} and \eqref{commdagger} look as follows \cite{Smolyakov:2021vlo}:
\begin{align}\label{commcoeff1}
&Y^{(1)}_{j,l}=\frac{\tilde M_{j,l}^{\scriptscriptstyle+}+\tilde M_{l,j}^{\scriptscriptstyle+}}{2}d_{l}-\frac{\tilde M_{j,-j-l}^{\scriptscriptstyle+}+\tilde M_{-j-l,j}^{\scriptscriptstyle+}}{2}d_{j+l}+\frac{\tilde M_{j,-l}^{\scriptscriptstyle-}+\tilde N_{-l,j}^{\scriptscriptstyle-}}{2}c_{l}-\frac{\tilde M_{j,j+l}^{\scriptscriptstyle-}+\tilde N_{j+l,j}^{\scriptscriptstyle-}}{2}c_{j+l},\\\label{commcoeff2}
&Y^{(2)}_{j,l}=\frac{\tilde N_{j,l}^{\scriptscriptstyle+}+\tilde N_{l,j}^{\scriptscriptstyle+}}{2}c_{l}-\frac{\tilde N_{j,-j-l}^{\scriptscriptstyle+}+\tilde N_{-j-l,j}^{\scriptscriptstyle+}}{2}c_{j+l}+\frac{\tilde N_{j,-l}^{\scriptscriptstyle-}+\tilde M_{-l,j}^{\scriptscriptstyle-}}{2}d_{l}-\frac{\tilde N_{j,j+l}^{\scriptscriptstyle-}+\tilde M_{j+l,j}^{\scriptscriptstyle-}}{2}d_{j+l},\\\label{commcoeff3}
&Y^{(3)}_{j,l}=\frac{\tilde M_{j,l}^{\scriptscriptstyle+}+\tilde M_{l,j}^{\scriptscriptstyle+}}{2}c_{l}+\frac{\tilde N_{j,-j-l}^{\scriptscriptstyle+}+\tilde N_{-j-l,j}^{\scriptscriptstyle+}}{2}d_{j+l}+\frac{\tilde M_{j,-l}^{\scriptscriptstyle-}+\tilde N_{-l,j}^{\scriptscriptstyle-}}{2}d_{l}+
\frac{\tilde N_{j,j+l}^{\scriptscriptstyle-}+\tilde M_{j+l,j}^{\scriptscriptstyle-}}{2}c_{j+l}.
\end{align}
It was shown in \cite{Smolyakov:2021vlo} that if $j$ and $l$ in \eqref{commcoeff1}--\eqref{commcoeff3} are such that for all terms in \eqref{commcoeff1}--\eqref{commcoeff3} relations \eqref{notransform1}, \eqref{notransform2} hold (namely, if $ \tilde M_{j,l}^{\scriptscriptstyle+}=M_{j,l}^{\star\scriptscriptstyle+}$, $\tilde N_{j,l}^{\scriptscriptstyle+}=N_{j,l}^{\star\scriptscriptstyle+}$ and $\tilde M_{j,l}^{\scriptscriptstyle-}=M_{j,l}^{\star\scriptscriptstyle-}$, $\tilde N_{j,l}^{\scriptscriptstyle-}=N_{j,l}^{\star\scriptscriptstyle-}$), these coefficients can be reduced to
\begin{align}\label{commcoeff1star}
&M_{j,l}^{\star\scriptscriptstyle+}d_{l}-M_{j,-j-l}^{\star\scriptscriptstyle+}d_{j+l}+M_{j,-l}^{\star\scriptscriptstyle-}c_{l}- M_{j,j+l}^{\star\scriptscriptstyle-}c_{j+l},\\\label{commcoeff2star}
&N_{j,l}^{\star\scriptscriptstyle+}c_{l}-N_{j,-j-l}^{\star\scriptscriptstyle+}c_{j+l}+N_{j,-l}^{\star\scriptscriptstyle-}d_{l}- N_{j,j+l}^{\star\scriptscriptstyle-}d_{j+l},\\\label{commcoeff3star}
&M_{j,l}^{\star\scriptscriptstyle+}c_{l}+N_{j,-j-l}^{\star\scriptscriptstyle+}d_{j+l}+M_{j,-l}^{\star\scriptscriptstyle-}d_{l}+
N_{j,j+l}^{\star\scriptscriptstyle-}c_{j+l}.
\end{align}
If $j$ and $l$ in \eqref{commcoeff1}--\eqref{commcoeff3} are such that at least for some of the terms in \eqref{commcoeff1}--\eqref{commcoeff3} relations \eqref{transform1}, \eqref{transform2} hold, then, as was also shown in \cite{Smolyakov:2021vlo}, with
\begin{align}\nonumber
Q_{a+b,a}^{\scriptscriptstyle-}=&
d_{b}\left(\frac{2}{c_{a+b}}\left(K_{a,b}^{\scriptscriptstyle+}d_{b}+M_{a,-b}^{\star\scriptscriptstyle-}c_{b}-M_{a,-a-b}^{\star\scriptscriptstyle+}d_{a+b}\right)
-K_{a,a+b}^{\scriptscriptstyle-}\right)\\\label{Qab+def}+&c_{b}
\left(\frac{2}{d_{a+b}}\left(N_{a,-b}^{\star\scriptscriptstyle-}d_{b}-N_{a,-a-b}^{\star\scriptscriptstyle+}c_{a+b}\right)
-K_{a+b,a}^{\scriptscriptstyle-}\right),\\\nonumber
Q_{a,a+b}^{\scriptscriptstyle-}=&
c_{b}\left(\frac{2}{c_{a+b}}\left(K_{a,b}^{\scriptscriptstyle+}d_{b}+M_{a,-b}^{\star\scriptscriptstyle-}c_{b}-M_{a,-a-b}^{\star\scriptscriptstyle+}d_{a+b}\right)
-K_{a,a+b}^{\scriptscriptstyle-}\right)\\\label{Qab-def}+&d_{b}
\left(\frac{2}{d_{a+b}}\left(N_{a,-b}^{\star\scriptscriptstyle-}d_{b}-N_{a,-a-b}^{\star\scriptscriptstyle+}c_{a+b}\right)
-K_{a+b,a}^{\scriptscriptstyle-}\right),
\end{align}
where $a$ and $b$ are such that $\gamma_{a+b}=\gamma_{a}+\gamma_{b}$, there remain only two independent coefficients in the corresponding commutation relations. These coefficients have the form
\begin{align}\label{commcoeff3secmod}
&K_{a,b}^{\scriptscriptstyle+}c_{b}-N_{a,-a-b}^{\star\scriptscriptstyle+}\frac{1}{d_{a+b}}
+N_{a,-b}^{\star\scriptscriptstyle-}\frac{d_{b}c_{a+b}}{d_{a+b}}+M_{a,-b}^{\star\scriptscriptstyle-}d_{b},\\\label{commcoeff4notstar}
&K_{a,b}^{\scriptscriptstyle+}\frac{d_{a+b}d_{b}}{c_{a+b}}+M_{a,-a-b}^{\star\scriptscriptstyle+}\frac{1}{c_{a+b}}
+N_{a,-b}^{\star\scriptscriptstyle-}c_{b}+M_{a,-b}^{\star\scriptscriptstyle-}\frac{d_{a+b}c_{b}}{c_{a+b}}.
\end{align}
In order not to overload the text, derivation of coefficients \eqref{commcoeff1star}--\eqref{commcoeff3star} and \eqref{commcoeff3secmod}, \eqref{commcoeff4notstar} is not reproduced in this paper --- detailed explanations and calculations can be found in \cite{Smolyakov:2021vlo}.
However, as now $M_{\scriptscriptstyle\#,\scriptscriptstyle\#}^{\star\scriptscriptstyle\pm}$, $N_{\scriptscriptstyle\#,\scriptscriptstyle\#}^{\star\scriptscriptstyle\pm}$, $K_{a,b}^{\scriptscriptstyle+}$, and $c_{\scriptscriptstyle\#}$, $d_{\scriptscriptstyle\#}$ depend on the parameters $\tilde v_{\scriptscriptstyle\#}$ and differ from those in \cite{Smolyakov:2021vlo}, coefficients \eqref{commcoeff1star}--\eqref{commcoeff3star} and \eqref{commcoeff3secmod}, \eqref{commcoeff4notstar} should be recalculated.

Exactly in the same way as it was done in \cite{Smolyakov:2021vlo}, let us express all terms with momenta in \eqref{commcoeff1star}--\eqref{commcoeff3star} through $\gamma_{\scriptscriptstyle\#}$ and $\tilde v_{\scriptscriptstyle\#}$. Using \eqref{cddef}, \eqref{qdef1}--\eqref{qdef4}, \eqref{coefftildeM+}, \eqref{coefftildeM-} and
{\small
\begin{equation}\label{relationsKtoGamma}
\frac{\vec k_{j}^{2}}{2m}=\sqrt{\gamma_{j}^{2}+(\omega\tilde v_{j})^{2}}-\omega\tilde v_{j},\,\,\frac{\vec k_{l}^{2}}{2m}=\sqrt{\gamma_{l}^{2}+(\omega\tilde v_{l})^{2}}-\omega\tilde v_{l},\,\,\frac{(\vec k_{j}+\vec k_{l})^{2}}{2m}=\sqrt{\gamma_{j+l}^{2}+(\omega\tilde v_{j+l})^{2}}-\omega\tilde v_{j+l},
\end{equation}}
the terms in the first coefficient \eqref{commcoeff1star} can be represented as
{\small
\begin{align}\nonumber
&M_{j,l}^{\star\scriptscriptstyle+}d_{l}=\frac{c_{j}}{2\tilde v_{j}\gamma_{l}\left(\left(\gamma_{j}+\gamma_{l}\right)^{2}-\gamma_{j+l}^{2}\right)}\Biggl[\left(\gamma_{j}+\gamma_{l}+\sqrt{\gamma_{j+l}^{2}+(\omega\tilde v_{j+l})^{2}}\right)\\\nonumber&\times\Biggl(\tilde v_{l}(\tilde v_{j}+\tilde v_{j+l})\left(\gamma_{j}-\sqrt{\gamma_{j}^{2}+(\omega\tilde v_{j})^{2}}\right)+\tilde v_{j}(\tilde v_{l}+\tilde v_{j+l})\left(\gamma_{l}
-\sqrt{\gamma_{l}^{2}+(\omega\tilde v_{l})^{2}}\right)+\tilde v_{j}\tilde v_{l}(\tilde v_{j}+\tilde v_{l})\omega\Biggr)\\\nonumber&-\tilde v_{j+l}(\tilde v_{j}+\tilde v_{l})\left(\gamma_{j}-\sqrt{\gamma_{j}^{2}+(\omega\tilde v_{j})^{2}}\right)
\left(\gamma_{l}-\sqrt{\gamma_{l}^{2}+(\omega\tilde v_{l})^{2}}\right)\\\label{termcomm1}&
-\omega\Biggl(\tilde v_{l}\tilde v_{j+l}(\tilde v_{l}+\tilde v_{j+l})\left(\gamma_{j}-\sqrt{\gamma_{j}^{2}+(\omega\tilde v_{j})^{2}}\right)+
\tilde v_{j}\tilde v_{j+l}(\tilde v_{j}+\tilde v_{j+l})\left(\gamma_{l}-\sqrt{\gamma_{l}^{2}+(\omega\tilde v_{l})^{2}}\right)\Biggr)\Biggr],
\end{align}
\begin{align}\nonumber
&M_{j,-j-l}^{\star\scriptscriptstyle+}d_{j+l}=\frac{c_{j}}{2\tilde v_{j}\gamma_{j+l}\left(\left(\gamma_{j}+\gamma_{j+l}\right)^{2}-\gamma_{l}^{2}\right)}\Biggl[\left(\gamma_{j}+\gamma_{j+l}+\sqrt{\gamma_{l}^{2}+(\omega\tilde v_{l})^{2}}\right)\\\nonumber&\times\Biggl(\tilde v_{j+l}(\tilde v_{j}+\tilde v_{l})\left(\gamma_{j}-\sqrt{\gamma_{j}^{2}+(\omega\tilde v_{j})^{2}}\right)+\tilde v_{j}(\tilde v_{l}+\tilde v_{j+l})\left(\gamma_{j+l}
-\sqrt{\gamma_{j+l}^{2}+(\omega\tilde v_{j+l})^{2}}\right)\\\nonumber&+\tilde v_{j}\tilde v_{j+l}(\tilde v_{j}+\tilde v_{j+l})\omega\Biggr)-\tilde v_{l}(\tilde v_{j}+\tilde v_{j+l})\left(\gamma_{j}-\sqrt{\gamma_{j}^{2}+(\omega\tilde v_{j})^{2}}\right)
\left(\gamma_{j+l}-\sqrt{\gamma_{j+l}^{2}+(\omega\tilde v_{j+l})^{2}}\right)\\\label{termcomm2}&
-\omega\Biggl(\tilde v_{l}\tilde v_{j+l}(\tilde v_{l}+\tilde v_{j+l})\left(\gamma_{j}-\sqrt{\gamma_{j}^{2}+(\omega\tilde v_{j})^{2}}\right)+
\tilde v_{j}\tilde v_{l}(\tilde v_{j}+\tilde v_{l})\left(\gamma_{j+l}-\sqrt{\gamma_{j+l}^{2}+(\omega\tilde v_{j+l})^{2}}\right)\Biggr)\Biggr],
\end{align}
\begin{align}\nonumber
&M_{j,-l}^{\star\scriptscriptstyle-}c_{l}=\frac{c_{j}}{2\tilde v_{j}\gamma_{l}\left(\left(\gamma_{j}-\gamma_{l}\right)^{2}-\gamma_{j+l}^{2}\right)}\Biggl[\left(\gamma_{j}-\gamma_{l}+\sqrt{\gamma_{j+l}^{2}+(\omega\tilde v_{j+l})^{2}}\right)\\\nonumber&\times\Biggl(\tilde v_{l}(\tilde v_{j}+\tilde v_{j+l})\left(\sqrt{\gamma_{j}^{2}+(\omega\tilde v_{j})^{2}}-\gamma_{j}\right)+\tilde v_{j}(\tilde v_{l}+\tilde v_{j+l})\left(\sqrt{\gamma_{l}^{2}+(\omega\tilde v_{l})^{2}}+\gamma_{l}\right)-\tilde v_{j}\tilde v_{l}(\tilde v_{j}+\tilde v_{l})\omega\Biggr)\\\nonumber&+\tilde v_{j+l}(\tilde v_{j}+\tilde v_{l})\left(\sqrt{\gamma_{j}^{2}+(\omega\tilde v_{j})^{2}}-\gamma_{j}\right)
\left(\sqrt{\gamma_{l}^{2}+(\omega\tilde v_{l})^{2}}+\gamma_{l}\right)\\\label{termcomm3}&
-\omega\Biggl(\tilde v_{l}\tilde v_{j+l}(\tilde v_{l}+\tilde v_{j+l})\left(\sqrt{\gamma_{j}^{2}+(\omega\tilde v_{j})^{2}}-\gamma_{j}\right)+
\tilde v_{j}\tilde v_{j+l}(\tilde v_{j}+\tilde v_{j+l})\left(\sqrt{\gamma_{l}^{2}+(\omega\tilde v_{l})^{2}}+\gamma_{l}\right)\Biggr)\Biggr],
\end{align}
\begin{align}\nonumber
&M_{j,j+l}^{\star\scriptscriptstyle-}c_{j+l}=\frac{c_{j}}{2\tilde v_{j}\gamma_{j+l}\left(\left(\gamma_{j}-\gamma_{j+l}\right)^{2}-\gamma_{l}^{2}\right)}\Biggl[\left(\gamma_{j}-\gamma_{j+l}+\sqrt{\gamma_{l}^{2}+(\omega\tilde v_{l})^{2}}\right)\\\nonumber&\times\Biggl(\tilde v_{j+l}(\tilde v_{j}+\tilde v_{l})\left(\sqrt{\gamma_{j}^{2}+(\omega\tilde v_{j})^{2}}-\gamma_{j}\right)+\tilde v_{j}(\tilde v_{l}+\tilde v_{j+l})\left(\sqrt{\gamma_{j+l}^{2}+(\omega\tilde v_{j+l})^{2}}+\gamma_{j+l}\right)\\\nonumber&-\tilde v_{j}\tilde v_{j+l}(\tilde v_{j}+\tilde v_{j+l})\omega\Biggr)+\tilde v_{l}(\tilde v_{j}+\tilde v_{j+l})\left(\sqrt{\gamma_{j}^{2}+(\omega\tilde v_{j})^{2}}-\gamma_{j}\right)
\left(\sqrt{\gamma_{j+l}^{2}+(\omega\tilde v_{j+l})^{2}}+\gamma_{j+l}\right)\\\label{termcomm4}&
-\omega\Biggl(\tilde v_{l}\tilde v_{j+l}(\tilde v_{l}+\tilde v_{j+l})\left(\sqrt{\gamma_{j}^{2}+(\omega\tilde v_{j})^{2}}-\gamma_{j}\right)+
\tilde v_{j}\tilde v_{l}(\tilde v_{j}+\tilde v_{l})\left(\sqrt{\gamma_{j+l}^{2}+(\omega\tilde v_{j+l})^{2}}+\gamma_{j+l}\right)\Biggr)\Biggr].
\end{align}}
One can show that
\begin{equation}\label{equality1zero}
M_{j,l}^{\star\scriptscriptstyle+}d_{l}-M_{j,-j-l}^{\star\scriptscriptstyle+}d_{j+l}+M_{j,-l}^{\star\scriptscriptstyle-}c_{l}- M_{j,j+l}^{\star\scriptscriptstyle-}c_{j+l}=0.
\end{equation}
It is a purely algebraic cancellation, i.e., equality \eqref{equality1zero} is valid for arbitrary nonzero values of $\gamma_{j}$, $\gamma_{l}$, $\gamma_{j+l}$ and $\tilde v_{j}$, $\tilde v_{l}$, $\tilde v_{j+l}$. Analytical check of equality \eqref{equality1zero} demands very bulky calculations, so it is more convenient to use a program package that can perform symbolic computations (namely, reduction and simplification of analytical expressions). To check the validity of equality \eqref{equality1zero} (and subsequent equalities of this section), the computer algebra system Maxima \cite{Maxima} was used, see Supplementary material for the files.

Analogously, for \eqref{commcoeff2star} one obtains
\begin{align}\nonumber
&N_{j,l}^{\star\scriptscriptstyle+}c_{l}-N_{j,-j-l}^{\star\scriptscriptstyle+}c_{j+l}+N_{j,-l}^{\star\scriptscriptstyle-}d_{l}- N_{j,j+l}^{\star\scriptscriptstyle-}d_{j+l}\\\nonumber
&=\frac{\gamma_{j}+\gamma_{l}-\sqrt{\gamma_{j+l}^{2}+(\omega\tilde v_{j+l})^{2}}}{\sqrt{\gamma_{l}^{2}+(\omega\tilde v_{l})^{2}}-\gamma_{l}}\cdot\frac{\tilde v_{l}}{\tilde v_{j+l}}\,M_{j,l}^{\star\scriptscriptstyle+}d_{l}
-\frac{\gamma_{j}+\gamma_{j+l}-\sqrt{\gamma_{l}^{2}+(\omega\tilde v_{l})^{2}}}{\sqrt{\gamma_{j+l}^{2}+(\omega\tilde v_{j+l})^{2}}-\gamma_{j+l}}\cdot\frac{\tilde v_{j+l}}{\tilde v_{l}}\, M_{j,-j-l}^{\star\scriptscriptstyle+}d_{j+l}\\\nonumber
&+\frac{\gamma_{j}-\gamma_{l}-\sqrt{\gamma_{j+l}^{2}+(\omega\tilde v_{j+l})^{2}}}{\sqrt{\gamma_{l}^{2}+(\omega\tilde v_{l})^{2}}+\gamma_{l}}\cdot\frac{\tilde v_{l}}{\tilde v_{j+l}}\, M_{j,-l}^{\star\scriptscriptstyle-}c_{l}-\frac{\gamma_{j}-\gamma_{j+l}-\sqrt{\gamma_{l}^{2}+(\omega\tilde v_{l})^{2}}}{\sqrt{\gamma_{j+l}^{2}+(\omega\tilde v_{j+l})^{2}}+\gamma_{j+l}}\cdot\frac{\tilde v_{j+l}}{\tilde v_{l}}\,M_{j,j+l}^{\star\scriptscriptstyle-}c_{j+l}\\\label{secondstarredcond}
&-\frac{1}{\omega}\left(\frac{q_{j,l}^{(1)}c_{l}}{\tilde v_{j+l}}-\frac{q_{j,-j-l}^{(1)}c_{j+l}}{\tilde v_{l}}+\frac{q_{j,-l}^{(3)}d_{l}}{\tilde v_{j+l}}-\frac{q_{j,j+l}^{(3)}d_{j+l}}{\tilde v_{l}}\right),
\end{align}
where the terms $M_{j,l}^{\star\scriptscriptstyle+}d_{l}$, $M_{j,-j-l}^{\star\scriptscriptstyle+}d_{j+l}$, $M_{j,-l}^{\star\scriptscriptstyle-}c_{l}$ and $M_{j,j+l}^{\star\scriptscriptstyle-}c_{j+l}$ are defined by \eqref{termcomm1}--\eqref{termcomm4}; Eqs. \eqref{tildeNeq+}, \eqref{tildeNeq-}, definitions \eqref{cddef}, \eqref{qdef1}--\eqref{qdef4} and relations \eqref{relationsKtoGamma} were used to pass from $N_{{\scriptscriptstyle\#},{\scriptscriptstyle\#}}^{\star\scriptscriptstyle\pm}$ to $M_{{\scriptscriptstyle\#},{\scriptscriptstyle\#}}^{\star\scriptscriptstyle\pm}$. Using the fact that $c_{j}^{2}-d_{j}^{2}=1$ for $j\neq 0$, for the terms in brackets in \eqref{secondstarredcond} one gets
\begin{align}\nonumber
&-\frac{1}{\omega}\left(\frac{q_{j,l}^{(1)}c_{l}}{\tilde v_{j+l}}-\frac{q_{j,-j-l}^{(1)}c_{j+l}}{\tilde v_{l}}+\frac{q_{j,-l}^{(3)}d_{l}}{\tilde v_{j+l}}-\frac{q_{j,j+l}^{(3)}d_{j+l}}{\tilde v_{l}}\right)\\&
=\frac{c_{j}}{\omega}\Biggl(\frac{\tilde v_{j}+\tilde v_{j+l}}{\tilde v_{l}}-\frac{\tilde v_{j}+\tilde v_{l}}{\tilde v_{j+l}}+\left(\frac{1}{\tilde v_{j+l}}-\frac{1}{\tilde v_{l}}\right)\frac{\sqrt{\gamma_{j}^{2}+(\omega\tilde v_{j})^{2}}-\gamma_{j}}{\omega}\Biggr),
\end{align}
leading to
\begin{align}\nonumber
&N_{j,l}^{\star\scriptscriptstyle+}c_{l}-N_{j,-j-l}^{\star\scriptscriptstyle+}c_{j+l}+N_{j,-l}^{\star\scriptscriptstyle-}d_{l}- N_{j,j+l}^{\star\scriptscriptstyle-}d_{j+l}\\\nonumber
&=\frac{\gamma_{j}+\gamma_{l}-\sqrt{\gamma_{j+l}^{2}+(\omega\tilde v_{j+l})^{2}}}{\sqrt{\gamma_{l}^{2}+(\omega\tilde v_{l})^{2}}-\gamma_{l}}\cdot\frac{\tilde v_{l}}{\tilde v_{j+l}}\,M_{j,l}^{\star\scriptscriptstyle+}d_{l}
-\frac{\gamma_{j}+\gamma_{j+l}-\sqrt{\gamma_{l}^{2}+(\omega\tilde v_{l})^{2}}}{\sqrt{\gamma_{j+l}^{2}+(\omega\tilde v_{j+l})^{2}}-\gamma_{j+l}}\cdot\frac{\tilde v_{j+l}}{\tilde v_{l}}\, M_{j,-j-l}^{\star\scriptscriptstyle+}d_{j+l}\\\nonumber
&+\frac{\gamma_{j}-\gamma_{l}-\sqrt{\gamma_{j+l}^{2}+(\omega\tilde v_{j+l})^{2}}}{\sqrt{\gamma_{l}^{2}+(\omega\tilde v_{l})^{2}}+\gamma_{l}}\cdot\frac{\tilde v_{l}}{\tilde v_{j+l}}\, M_{j,-l}^{\star\scriptscriptstyle-}c_{l}-\frac{\gamma_{j}-\gamma_{j+l}-\sqrt{\gamma_{l}^{2}+(\omega\tilde v_{l})^{2}}}{\sqrt{\gamma_{j+l}^{2}+(\omega\tilde v_{j+l})^{2}}+\gamma_{j+l}}\cdot\frac{\tilde v_{j+l}}{\tilde v_{l}}\,M_{j,j+l}^{\star\scriptscriptstyle-}c_{j+l}\\\label{secondstarredcondfinal}
&+\frac{c_{j}}{\omega}\Biggl(\frac{\tilde v_{j}+\tilde v_{j+l}}{\tilde v_{l}}-\frac{\tilde v_{j}+\tilde v_{l}}{\tilde v_{j+l}}+\left(\frac{1}{\tilde v_{j+l}}-\frac{1}{\tilde v_{l}}\right)\frac{\sqrt{\gamma_{j}^{2}+(\omega\tilde v_{j})^{2}}-\gamma_{j}}{\omega}\Biggr).
\end{align}
One can show that
\begin{equation}\label{equality2zero}
N_{j,l}^{\star\scriptscriptstyle+}c_{l}-N_{j,-j-l}^{\star\scriptscriptstyle+}c_{j+l}+N_{j,-l}^{\star\scriptscriptstyle-}d_{l}- N_{j,j+l}^{\star\scriptscriptstyle-}d_{j+l}=0
\end{equation}
for arbitrary nonzero values of $\gamma_{j}$, $\gamma_{l}$, $\gamma_{j+l}$ and $\tilde v_{j}$, $\tilde v_{l}$, $\tilde v_{j+l}$, see Supplementary material.

As for coefficient \eqref{commcoeff3star}, with the help of Eqs.~\eqref{tildeNeq+}, \eqref{tildeNeq-} it can be brought to the form
\begin{align}\nonumber
&M_{j,l}^{\star\scriptscriptstyle+}c_{l}+N_{j,-j-l}^{\star\scriptscriptstyle+}d_{j+l}+M_{j,-l}^{\star\scriptscriptstyle-}d_{l}+
N_{j,j+l}^{\star\scriptscriptstyle-}c_{j+l}=\frac{\sqrt{\gamma_{l}^{2}+(\omega\tilde v_{l})^{2}}+\gamma_{l}}{\omega\tilde v_{l}}\,M_{j,l}^{\star\scriptscriptstyle+}d_{l}\\\nonumber
&+\frac{\gamma_{j}+\gamma_{j+l}-\sqrt{\gamma_{l}^{2}+(\omega\tilde v_{l})^{2}}}{\omega\tilde v_{l}}\,M_{j,-j-l}^{\star\scriptscriptstyle+}d_{j+l}+\frac{\sqrt{\gamma_{l}^{2}+(\omega\tilde v_{l})^{2}}-\gamma_{l}}{\omega\tilde v_{l}}\,M_{j,-l}^{\star\scriptscriptstyle-}c_{l}\\\nonumber
&+\frac{\gamma_{j}-\gamma_{j+l}-\sqrt{\gamma_{l}^{2}+(\omega\tilde v_{l})^{2}}}{\omega\tilde v_{l}}\,M_{j,j+l}^{\star\scriptscriptstyle-}c_{j+l}
-\frac{1}{\omega\tilde v_{l}}\left(q_{j,-j-l}^{(1)}d_{j+l}+q_{j,j+l}^{(3)}c_{j+l}\right)\\
&=\frac{1}{\omega\tilde v_{l}}\Bigl(\gamma_{l}M_{j,l}^{\star\scriptscriptstyle+}d_{l}
+(\gamma_{j}+\gamma_{j+l})M_{j,-j-l}^{\star\scriptscriptstyle+}d_{j+l}-\gamma_{l}M_{j,-l}^{\star\scriptscriptstyle-}c_{l}
+(\gamma_{j}-\gamma_{j+l})M_{j,j+l}^{\star\scriptscriptstyle-}c_{j+l}-c_{j}(\tilde v_{l}+\tilde v_{j+l})\Bigr),
\end{align}
where the terms $M_{j,l}^{\star\scriptscriptstyle+}d_{l}$, $M_{j,-j-l}^{\star\scriptscriptstyle+}d_{j+l}$, $M_{j,-l}^{\star\scriptscriptstyle-}c_{l}$ and $M_{j,j+l}^{\star\scriptscriptstyle-}c_{j+l}$ are defined by \eqref{termcomm1}--\eqref{termcomm4}, and definitions \eqref{cddef}, \eqref{qdef1}--\eqref{qdef4} and Eq.~\eqref{equality1zero} were used. One can also show that
\begin{equation}\label{equality3zero}
M_{j,l}^{\star\scriptscriptstyle+}c_{l}+N_{j,-j-l}^{\star\scriptscriptstyle+}d_{j+l}+M_{j,-l}^{\star\scriptscriptstyle-}d_{l}+
N_{j,j+l}^{\star\scriptscriptstyle-}c_{j+l}=0
\end{equation}
for arbitrary nonzero values of $\gamma_{j}$, $\gamma_{l}$, $\gamma_{j+l}$ and $\tilde v_{j}$, $\tilde v_{l}$, $\tilde v_{j+l}$, see Supplementary material.

Now let us turn to expression \eqref{commcoeff3secmod}. Again, exactly as it was done in \cite{Smolyakov:2021vlo}, it is necessary to perform the following three steps:\\
{\bf 1.} With the help of Eqs.~\eqref{tildeNeq+} and \eqref{tildeNeq-}, one should pass from $N_{a,-a-b}^{\star\scriptscriptstyle+}$ to $M_{a,-a-b}^{\star\scriptscriptstyle+}$ and from $N_{a,-b}^{\star\scriptscriptstyle-}$ to $M_{a,-b}^{\star\scriptscriptstyle-}$.\\
{\bf 2.} With the help of \eqref{cddef}, \eqref{qdef1}--\eqref{qdef4}, and \eqref{relationsKtoGamma}, one should express all terms with momenta in \eqref{commcoeff3secmod} through $\gamma_{\scriptscriptstyle\#}$ and $\tilde v_{\scriptscriptstyle\#}$.\\
{\bf 3.} One should replace $\gamma_{a+b}$ by $\gamma_{a}+\gamma_{b}$ in all places where $\gamma_{a+b}$ emerges.\\
The result is
\begin{align}\nonumber
&K_{a,b}^{\scriptscriptstyle+}c_{b}-N_{a,-a-b}^{\star\scriptscriptstyle+}\frac{1}{d_{a+b}}
+N_{a,-b}^{\star\scriptscriptstyle-}\frac{d_{b}c_{a+b}}{d_{a+b}}+M_{a,-b}^{\star\scriptscriptstyle-}d_{b}\\\nonumber
&=K_{a,b}^{\scriptscriptstyle+}c_{b}-\frac{2}
{\omega\tilde v_{b}\left(\sqrt{\left(\gamma_{a}+\gamma_{b}\right)^{2}+(\omega\tilde v_{a+b})^{2}}-\gamma_{a}-\gamma_{b}\right)}\\\nonumber&\times\left[
\left(\gamma_{a}+\gamma_{b}\right)\left(2\gamma_{a}+\gamma_{b}-\sqrt{\gamma_{b}^{2}+(\omega\tilde v_{b})^{2}}\right)M_{a,-a-b}^{\star\scriptscriptstyle+}d_{a+b}
+\gamma_{b}\left(\sqrt{\gamma_{b}^{2}+(\omega\tilde v_{b})^{2}}-\gamma_{b}\right)M_{a,-b}^{\star\scriptscriptstyle-}c_{b}\right]
\\\nonumber&+\frac{c_{a}}{2\,\tilde v_{a}\tilde v_{b}\left(\sqrt{\left(\gamma_{a}+\gamma_{b}\right)^{2}+(\omega\tilde v_{a+b})^{2}}-\gamma_{a}-\gamma_{b}\right)}\Biggl[2\,\tilde v_{a}\tilde v_{a+b}(\tilde v_{a}+\tilde v_{a+b})\\\nonumber&
-2\,\tilde v_{a}(\tilde v_{b}+\tilde v_{a+b})\frac{\sqrt{\left(\gamma_{a}+\gamma_{b}\right)^{2}+(\omega\tilde v_{a+b})^{2}}-\gamma_{a}-\gamma_{b}}{\omega}
-2\,\tilde v_{a+b}(\tilde v_{a}+\tilde v_{b})\frac{\sqrt{\gamma_{a}^{2}+(\omega\tilde v_{a})^{2}}-\gamma_{a}}{\omega}
\\\nonumber&-\tilde v_{b}(\tilde v_{a}+\tilde v_{a+b})\frac{\left(\sqrt{\gamma_{a}^{2}+(\omega\tilde v_{a})^{2}}-\gamma_{a}\right)
\left(\sqrt{\gamma_{b}^{2}+(\omega\tilde v_{b})^{2}}-\gamma_{b}\right)}{\omega\gamma_{b}}-\tilde v_{a}{\tilde v_{b}}^{2}(\tilde v_{b}+\tilde v_{a+b})\frac{\omega}{\gamma_{b}}\\&+\tilde v_{a}\tilde v_{b}(\tilde v_{a}+\tilde v_{b})\frac{\sqrt{\gamma_{b}^{2}+(\omega\tilde v_{b})^{2}}-\gamma_{b}}{\gamma_{b}}\Biggr],
\end{align}
where $M_{a,-a-b}^{\star\scriptscriptstyle+}d_{a+b}$ is defined by \eqref{termcomm2}, $M_{a,-b}^{\star\scriptscriptstyle-}c_{b}$ is defined by \eqref{termcomm3}, $\gamma_{a+b}$ replaced by $\gamma_{a}+\gamma_{b}$, and
\begin{align}\nonumber
&K_{a,b}^{\scriptscriptstyle+}c_{b}
=\frac{c_{a}\omega}{4\tilde v_{a}\tilde v_{a+b}\gamma_{b}\left(\gamma_{a}+\gamma_{b}\right)}\Biggl[
\frac{\sqrt{\left(\gamma_{a}+\gamma_{b}\right)^{2}+(\omega\tilde v_{a+b})^{2}}+\gamma_{a}+\gamma_{b}}{\omega}\\\nonumber&\times
\Biggl((\tilde v_{b}+\tilde v_{a+b})\frac{\left(\sqrt{\gamma_{a}^{2}+(\omega\tilde v_{a})^{2}}-\gamma_{a}\right)\left(\sqrt{\gamma_{b}^{2}+(\omega\tilde v_{b})^{2}}+\gamma_{b}\right)}{\omega^{2}}-\tilde v_{b}(\tilde v_{a}+\tilde v_{b})\frac{\sqrt{\gamma_{a}^{2}+(\omega\tilde v_{a})^{2}}-\gamma_{a}}{\omega}\\\nonumber&+\tilde v_{a}\tilde v_{b}(\tilde v_{a}+\tilde v_{a+b})\Biggr)
+\tilde v_{a}\tilde v_{a+b}(\tilde v_{a}+\tilde v_{b})\frac{\sqrt{\gamma_{b}^{2}+(\omega\tilde v_{b})^{2}}+\gamma_{b}}{\omega}
-\tilde v_{a}\tilde v_{b}\tilde v_{a+b}(\tilde v_{b}+\tilde v_{a+b})\\\label{K+cbdef}
&-\tilde v_{a+b}(\tilde v_{a}+\tilde v_{a+b})
\frac{\left(\sqrt{\gamma_{a}^{2}+(\omega\tilde v_{a})^{2}}-\gamma_{a}\right)\left(\sqrt{\gamma_{b}^{2}+(\omega\tilde v_{b})^{2}}+\gamma_{b}\right)}{\omega^{2}}\Biggr].
\end{align}
One can show that
\begin{equation}\label{equality4zero}
K_{a,b}^{\scriptscriptstyle+}c_{b}-N_{a,-a-b}^{\star\scriptscriptstyle+}\frac{1}{d_{a+b}}
+N_{a,-b}^{\star\scriptscriptstyle-}\frac{d_{b}c_{a+b}}{d_{a+b}}+M_{a,-b}^{\star\scriptscriptstyle-}d_{b}=0
\end{equation}
for arbitrary nonzero values of $\gamma_{a}$, $\gamma_{b}$ and $\tilde v_{a}$, $\tilde v_{b}$, $\tilde v_{a+b}$, see Supplementary material.

Fully analogous calculations can be made for \eqref{commcoeff4notstar}, leading to
\begin{align}\nonumber
&K_{a,b}^{\scriptscriptstyle+}\frac{d_{a+b}d_{b}}{c_{a+b}}+M_{a,-a-b}^{\star\scriptscriptstyle+}\frac{1}{c_{a+b}}
+N_{a,-b}^{\star\scriptscriptstyle-}c_{b}+M_{a,-b}^{\star\scriptscriptstyle-}\frac{d_{a+b}c_{b}}{c_{a+b}}\\\nonumber
=&\frac{\sqrt{\left(\gamma_{a}+\gamma_{b}\right)^{2}+(\omega\tilde v_{a+b})^{2}}-\gamma_{a}-\gamma_{b}}
{\sqrt{\gamma_{b}^{2}+(\omega\tilde v_{b})^{2}}+\gamma_{b}}\cdot\frac{\tilde v_{b}}{\tilde v_{a+b}}K_{a,b}^{\scriptscriptstyle+}c_{b}+\frac{2}{\omega\tilde v_{a+b}}\left(\left(\gamma_{a}+\gamma_{b}\right)M_{a,-a-b}^{\star\scriptscriptstyle+}d_{a+b}
-\gamma_{b}M_{a,-b}^{\star\scriptscriptstyle-}c_{b}\right)\\\nonumber&
-\frac{c_{a}}{2\tilde v_{a}\tilde v_{a+b}\gamma_{b}}\Biggl(\tilde v_{b}(\tilde v_{a}+\tilde v_{a+b})\frac{\sqrt{\gamma_{a}^{2}+(\omega\tilde v_{a})^{2}}-\gamma_{a}}{\omega}+\tilde v_{a}(\tilde v_{b}+\tilde v_{a+b})\frac{\sqrt{\gamma_{b}^{2}+(\omega\tilde v_{b})^{2}}+\gamma_{b}}{\omega}
\\&-\tilde v_{a}\tilde v_{b}(\tilde v_{a}+\tilde v_{b})\Biggr),
\end{align}
where $M_{a,-a-b}^{\star\scriptscriptstyle+}d_{a+b}$ is defined by \eqref{termcomm2}, $M_{a,-b}^{\star\scriptscriptstyle-}c_{b}$ is defined by \eqref{termcomm3}, $\gamma_{a+b}$ replaced by $\gamma_{a}+\gamma_{b}$, and $K_{a,b}^{\scriptscriptstyle+}c_{b}$ is defined by \eqref{K+cbdef}. Analogously, one can show that
\begin{equation}\label{equality5zero}
K_{a,b}^{\scriptscriptstyle+}\frac{d_{a+b}d_{b}}{c_{a+b}}+M_{a,-a-b}^{\star\scriptscriptstyle+}\frac{1}{c_{a+b}}
+N_{a,-b}^{\star\scriptscriptstyle-}c_{b}+M_{a,-b}^{\star\scriptscriptstyle-}\frac{d_{a+b}c_{b}}{c_{a+b}}=0
\end{equation}
for arbitrary nonzero values of $\gamma_{a}$, $\gamma_{b}$ and $\tilde v_{a}$, $\tilde v_{b}$, $\tilde v_{a+b}$, see Supplementary material.

Thus, all five coefficients \eqref{commcoeff1star}--\eqref{commcoeff3star} and \eqref{commcoeff3secmod}, \eqref{commcoeff4notstar} are equal to zero, which means that the operator $\hat{\tilde\phi}(t,\vec x)$ satisfies commutation relations \eqref{CCR1} and \eqref{CCR2}.

\subsubsection{Commutation relations for the operators $\hat\rho_{j,l}^{\scriptscriptstyle\pm}(t,\vec x)$}\label{subsubsectrho}
It was explained in detail in \cite{Smolyakov:2021vlo} that in order to correctly calculate contributions of the terms with $\hat\rho_{j,l}^{\scriptscriptstyle\pm}(t,\vec x)$ to the canonical commutation relations, it is necessary to consider the combination
\begin{equation}\label{rhocombin}
\hat\rho_{a,b}^{\scriptscriptstyle+}(t,\vec x)+\hat\rho_{a+b,b}^{\scriptscriptstyle-}(t,\vec x)+\hat\rho_{a+b,a}^{\scriptscriptstyle-}(t,\vec x),
\end{equation}
where $a$ and $b$ are such that $\gamma_{a+b}=\gamma_{a}+\gamma_{b}$. Since in terms of $L_{j,l}^{\scriptscriptstyle\pm}$ and $J_{j,l}^{\scriptscriptstyle\pm}$ the operators $\hat\rho_{j,l}^{\scriptscriptstyle\pm}(t,\vec x)$ defined by \eqref{rho+} and \eqref{rho-} have exactly the same form as those in \cite{Smolyakov:2021vlo}, again we can use the explicit expressions for the corresponding commutators obtained in \cite{Smolyakov:2021vlo}. The first commutation relation for combination \eqref{rhocombin} has the form \cite{Smolyakov:2021vlo}
\begin{align}\nonumber
&[\hat\rho_{a,b}^{\scriptscriptstyle+}(t,\vec x)+\hat\rho_{a+b,b}^{\scriptscriptstyle-}(t,\vec x)+\hat\rho_{a+b,a}^{\scriptscriptstyle-}(t,\vec x),\hat\varphi(t,\vec y)]+[\hat\varphi(t,\vec x),\hat\rho_{a,b}^{\scriptscriptstyle+}(t,\vec y)+\hat\rho_{a+b,b}^{\scriptscriptstyle-}(t,\vec y)+\hat\rho_{a+b,a}^{\scriptscriptstyle-}(t,\vec y)]\\\nonumber
&=-\frac{\omega t}{L^{3}}\Biggl(\Bigl(e^{\frac{i}{\hbar}(\vec k_{a}+\vec k_{b})\vec x}e^{-\frac{i}{\hbar}\vec k_{b}\vec y}e^{-\frac{i}{\hbar}\gamma_{a}t}\hat a_{a}^{}\left(L_{a,b}^{\scriptscriptstyle+}d_{b}-J_{a+b,a}^{\scriptscriptstyle-}c_{a+b}\right)\\\nonumber&+
e^{-\frac{i}{\hbar}(\vec k_{a}+\vec k_{b})\vec x}e^{\frac{i}{\hbar}\vec k_{b}\vec y}e^{\frac{i}{\hbar}\gamma_{a}t}\hat a_{a}^{\dagger}\left(J_{a,b}^{\scriptscriptstyle+}c_{b}-L_{a+b,a}^{\scriptscriptstyle-}d_{a+b}\right)+\left[a\leftrightarrow b\right]\Bigr)\\\nonumber
&+e^{\frac{i}{\hbar}\vec k_{a}\vec x}e^{\frac{i}{\hbar}\vec k_{b}\vec y}e^{-\frac{i}{\hbar}(\gamma_{a}+\gamma_{b})t}\hat a_{a+b}^{}\left(L_{a+b,b}^{\scriptscriptstyle-}c_{b}-L_{a+b,a}^{\scriptscriptstyle-}c_{a}\right)\\\label{commTterms1}
&+e^{-\frac{i}{\hbar}\vec k_{a}\vec x}e^{-\frac{i}{\hbar}\vec k_{b}\vec y}e^{\frac{i}{\hbar}(\gamma_{a}+\gamma_{b})t}\hat a_{a+b}^{\dagger}\left(J_{a+b,b}^{\scriptscriptstyle-}d_{b}-J_{a+b,a}^{\scriptscriptstyle-}d_{a}\right)
-\left[\vec x\leftrightarrow\vec y\right]\Biggr).
\end{align}
Recall that if $j$ and $l$ are such that relation \eqref{transform1} holds for these $j$ and $l$, the relation
\begin{equation}\label{JLtrs1}
J_{j,l}^{\scriptscriptstyle+}=\frac{\sqrt{\left(\gamma_{j}+\gamma_{l}\right)^{2}+(\omega\tilde v_{j+l})^{2}}-(\gamma_{j}+\gamma_{l})}{\omega\tilde v_{j+l}}L_{j,l}^{\scriptscriptstyle+},
\end{equation}
following from Eq.~\eqref{LJeq+} and solutions \eqref{L+def}, \eqref{J+def}, is also valid. Analogously, if $j$ and $l$ are such that relation \eqref{transform2} holds, the relation
\begin{equation}\label{JLtrs2}
J_{j,l}^{\scriptscriptstyle-}=\frac{\sqrt{\left(\gamma_{j}-\gamma_{l}\right)^{2}+(\omega\tilde v_{j-l})^{2}}-(\gamma_{j}-\gamma_{l})}{\omega\tilde v_{j-l}}L_{j,l}^{\scriptscriptstyle-},
\end{equation}
following from Eq.~\eqref{LJeq-} and solutions \eqref{L-def}, \eqref{J-def}, is also valid. Using formulas \eqref{JLtrs1} and \eqref{JLtrs2}, commutators \eqref{commTterms1} can be represented as
\begin{align}\nonumber
&[\hat\rho_{a,b}^{\scriptscriptstyle+}(t,\vec x)+\hat\rho_{a+b,b}^{\scriptscriptstyle-}(t,\vec x)+\hat\rho_{a+b,a}^{\scriptscriptstyle-}(t,\vec x),\hat\varphi(t,\vec y)]+[\hat\varphi(t,\vec x),\hat\rho_{a,b}^{\scriptscriptstyle+}(t,\vec y)+\hat\rho_{a+b,b}^{\scriptscriptstyle-}(t,\vec y)+\hat\rho_{a+b,a}^{\scriptscriptstyle-}(t,\vec y)]\\\nonumber
&=-\frac{\omega t}{L^{3}}\Biggl(\Biggl(e^{\frac{i}{\hbar}(\vec k_{a}+\vec k_{b})\vec x}e^{-\frac{i}{\hbar}\vec k_{b}\vec y}e^{-\frac{i}{\hbar}\gamma_{a}t}\hat a_{a}^{}\frac{\sqrt{\gamma_{b}^{2}+(\omega\tilde v_{b})^{2}}-\gamma_{b}}{\omega\tilde v_{b}}
\left(L_{a,b}^{\scriptscriptstyle+}c_{b}-L_{a+b,a}^{\scriptscriptstyle-}c_{a+b}\right)\\\nonumber&+
e^{-\frac{i}{\hbar}(\vec k_{a}+\vec k_{b})\vec x}e^{\frac{i}{\hbar}\vec k_{b}\vec y}e^{\frac{i}{\hbar}\gamma_{a}t}\hat a_{a}^{\dagger}
\frac{\sqrt{\left(\gamma_{a}+\gamma_{b}\right)^{2}+(\omega\tilde v_{a+b})^{2}}-(\gamma_{a}+\gamma_{b})}{\omega\tilde v_{a+b}}
\left(L_{a,b}^{\scriptscriptstyle+}c_{b}-L_{a+b,a}^{\scriptscriptstyle-}c_{a+b}\right)\\\nonumber
&+\left[a\leftrightarrow b\right]\Biggr)+e^{\frac{i}{\hbar}\vec k_{a}\vec x}e^{\frac{i}{\hbar}\vec k_{b}\vec y}e^{-\frac{i}{\hbar}(\gamma_{a}+\gamma_{b})t}\hat a_{a+b}^{}\left(L_{a+b,b}^{\scriptscriptstyle-}c_{b}-L_{a+b,a}^{\scriptscriptstyle-}c_{a}\right)
\\\label{commTterms2}
&+e^{-\frac{i}{\hbar}\vec k_{a}\vec x}e^{-\frac{i}{\hbar}\vec k_{b}\vec y}e^{\frac{i}{\hbar}(\gamma_{a}+\gamma_{b})t}\hat a_{a+b}^{\dagger}
\frac{\sqrt{\gamma_{a}^{2}+(\omega\tilde v_{a})^{2}}-\gamma_{a}}
{\sqrt{\gamma_{b}^{2}+(\omega\tilde v_{b})^{2}}+\gamma_{b}}\left(L_{a+b,b}^{\scriptscriptstyle-}c_{b}-L_{a+b,a}^{\scriptscriptstyle-}c_{a}\right)
-\left[\vec x\leftrightarrow\vec y\right]\Biggr).
\end{align}
The remaining independent coefficients in \eqref{commTterms2} are just
\begin{align}\label{Lcoeff1}
&L_{a,b}^{\scriptscriptstyle+}c_{b}-L_{a+b,a}^{\scriptscriptstyle-}c_{a+b},\\\label{Lcoeff2}
&L_{a+b,b}^{\scriptscriptstyle-}c_{b}-L_{a+b,a}^{\scriptscriptstyle-}c_{a}.
\end{align}
Note that in terms of $L_{\scriptscriptstyle\#,\scriptscriptstyle\#}^{\scriptscriptstyle\pm}$ and $c_{\scriptscriptstyle\#}$ they have the same form as those in \cite{Smolyakov:2021vlo}.

The second commutation relation for combination \eqref{rhocombin} has the form \cite{Smolyakov:2021vlo}
\begin{align}\nonumber
&[\hat\rho_{a,b}^{\scriptscriptstyle+}(t,\vec x)+\hat\rho_{a+b,b}^{\scriptscriptstyle-}(t,\vec x)+\hat\rho_{a+b,a}^{\scriptscriptstyle-}(t,\vec x),\hat\varphi^{\dagger}(t,\vec y)]+[\hat\varphi(t,\vec x),\left(\hat\rho_{a,b}^{\scriptscriptstyle+}(t,\vec y)+\hat\rho_{a+b,b}^{\scriptscriptstyle-}(t,\vec y)+\hat\rho_{a+b,a}^{\scriptscriptstyle-}(t,\vec y)\right)^{\dagger}]\\\nonumber
&=\frac{\omega t}{L^{3}}\Biggl(\Bigl(e^{\frac{i}{\hbar}(\vec k_{a}+\vec k_{b})\vec x}e^{-\frac{i}{\hbar}\vec k_{b}\vec y}e^{-\frac{i}{\hbar}\gamma_{a}t}\hat a_{a}^{}\left(L_{a,b}^{\scriptscriptstyle+}c_{b}-L_{a+b,a}^{\scriptscriptstyle-}c_{a+b}\right)\\\nonumber&+
e^{-\frac{i}{\hbar}(\vec k_{a}+\vec k_{b})\vec x}e^{\frac{i}{\hbar}\vec k_{b}\vec y}e^{\frac{i}{\hbar}\gamma_{a}t}\hat a_{a}^{\dagger}\left(J_{a,b}^{\scriptscriptstyle+}d_{b}-J_{a+b,a}^{\scriptscriptstyle-}d_{a+b}\right)+\left[a\leftrightarrow b\right]\Bigr)\\\nonumber
&+e^{\frac{i}{\hbar}\vec k_{a}\vec x}e^{\frac{i}{\hbar}\vec k_{b}\vec y}e^{-\frac{i}{\hbar}(\gamma_{a}+\gamma_{b})t}\hat a_{a+b}^{}\left(L_{a+b,b}^{\scriptscriptstyle-}d_{b}-J_{a+b,a}^{\scriptscriptstyle-}c_{a}\right)\\\label{commTterms3}
&+e^{-\frac{i}{\hbar}\vec k_{a}\vec x}e^{-\frac{i}{\hbar}\vec k_{b}\vec y}e^{\frac{i}{\hbar}(\gamma_{a}+\gamma_{b})t}\hat a_{a+b}^{\dagger}\left(J_{a+b,b}^{\scriptscriptstyle-}c_{b}-L_{a+b,a}^{\scriptscriptstyle-}d_{a}\right)
+\left[\vec x\leftrightarrow\vec y\right]^{\dagger}\Biggr).
\end{align}
Using relations \eqref{JLtrs1} and \eqref{JLtrs2}, commutators in \eqref{commTterms3} can be represented as
\begin{align}\nonumber
&[\hat\rho_{a,b}^{\scriptscriptstyle+}(t,\vec x)+\hat\rho_{a+b,b}^{\scriptscriptstyle-}(t,\vec x)+\hat\rho_{a+b,a}^{\scriptscriptstyle-}(t,\vec x),\hat\varphi^{\dagger}(t,\vec y)]+[\hat\varphi(t,\vec x),\left(\hat\rho_{a,b}^{\scriptscriptstyle+}(t,\vec y)+\hat\rho_{a+b,b}^{\scriptscriptstyle-}(t,\vec y)+\hat\rho_{a+b,a}^{\scriptscriptstyle-}(t,\vec y)\right)^{\dagger}]\\\nonumber
&=\frac{\omega t}{L^{3}}\Biggl(\Biggl(e^{\frac{i}{\hbar}(\vec k_{a}+\vec k_{b})\vec x}e^{-\frac{i}{\hbar}\vec k_{b}\vec y}e^{-\frac{i}{\hbar}\gamma_{a}t}\hat a_{a}^{}\left(L_{a,b}^{\scriptscriptstyle+}c_{b}-L_{a+b,a}^{\scriptscriptstyle-}c_{a+b}\right)\\\nonumber&+
e^{-\frac{i}{\hbar}(\vec k_{a}+\vec k_{b})\vec x}e^{\frac{i}{\hbar}\vec k_{b}\vec y}e^{\frac{i}{\hbar}\gamma_{a}t}\hat a_{a}^{\dagger}\frac{\sqrt{\left(\gamma_{a}+\gamma_{b}\right)^{2}+(\omega\tilde v_{a+b})^{2}}-(\gamma_{a}+\gamma_{b})}
{\sqrt{\gamma_{b}^{2}+(\omega\tilde v_{b})^{2}}+\gamma_{b}}\left(L_{a,b}^{\scriptscriptstyle+}c_{b}-L_{a+b,a}^{\scriptscriptstyle-}c_{a+b}\right)\\\nonumber
&+\left[a\leftrightarrow b\right]\Biggr)+e^{\frac{i}{\hbar}\vec k_{a}\vec x}e^{\frac{i}{\hbar}\vec k_{b}\vec y}e^{-\frac{i}{\hbar}(\gamma_{a}+\gamma_{b})t}\hat a_{a+b}^{}\frac{\sqrt{\gamma_{b}^{2}+(\omega\tilde v_{b})^{2}}-\gamma_{b}}{\omega\tilde v_{b}}\left(L_{a+b,b}^{\scriptscriptstyle-}c_{b}
-L_{a+b,a}^{\scriptscriptstyle-}c_{a}\right)\\\label{commTterms4}
&+e^{-\frac{i}{\hbar}\vec k_{a}\vec x}e^{-\frac{i}{\hbar}\vec k_{b}\vec y}e^{\frac{i}{\hbar}(\gamma_{a}+\gamma_{b})t}\hat a_{a+b}^{\dagger}\frac{\sqrt{\gamma_{a}^{2}+(\omega\tilde v_{a})^{2}}-\gamma_{a}}{\omega\tilde v_{a}}
\left(L_{a+b,b}^{\scriptscriptstyle-}c_{b}-L_{a+b,a}^{\scriptscriptstyle-}c_{a}\right)
+\left[\vec x\leftrightarrow\vec y\right]^{\dagger}\Biggr).
\end{align}
Here the remaining independent coefficients are \eqref{Lcoeff1} and \eqref{Lcoeff2} as well.

However, as now $L_{\scriptscriptstyle\#,\scriptscriptstyle\#}^{\scriptscriptstyle\pm}$, $J_{\scriptscriptstyle\#,\scriptscriptstyle\#}^{\scriptscriptstyle\pm}$ and $c_{\scriptscriptstyle\#}$, $d_{\scriptscriptstyle\#}$ depend on the parameters $\tilde v_{\scriptscriptstyle\#}$ and differ from those in \cite{Smolyakov:2021vlo}, coefficients in \eqref{commTterms1} should be recalculated. Using \eqref{cddef}, \eqref{qdef1}--\eqref{qdef4}, \eqref{L+def}, \eqref{L-def}, and
\begin{align}
&\frac{\vec k_{a}^{2}}{2m}=\sqrt{\gamma_{a}^{2}+(\omega\tilde v_{a})^{2}}-\omega\tilde v_{a},\qquad \frac{\vec k_{b}^{2}}{2m}=\sqrt{\gamma_{b}^{2}+(\omega\tilde v_{b})^{2}}-\omega\tilde v_{b},\\&\frac{(\vec k_{a}+\vec k_{b})^{2}}{2m}=\sqrt{\left(\gamma_{a}+\gamma_{b}\right)^{2}+(\omega\tilde v_{a+b})^{2}}-\omega\tilde v_{a+b},
\end{align}
it is not difficult to check that
\begin{align}\label{lastbut1eqcoeff}
&L_{a,b}^{\scriptscriptstyle+}c_{b}-L_{a+b,a}^{\scriptscriptstyle-}c_{a+b}=0,\\\label{lasteqcoeff}
&L_{a+b,b}^{\scriptscriptstyle-}c_{b}-L_{a+b,a}^{\scriptscriptstyle-}c_{a}=0
\end{align}
for arbitrary nonzero values of $\gamma_{a}$, $\gamma_{b}$ and $\tilde v_{a}$, $\tilde v_{b}$, $\tilde v_{a+b}$, see Appendix for details. Equalities \eqref{lastbut1eqcoeff} and \eqref{lasteqcoeff} imply that commutation relations \eqref{CCR1} and \eqref{CCR2} are also satisfied for the terms containing the operators $\hat\rho_{j,l}^{\scriptscriptstyle\pm}(t,\vec x)$.

\subsection{Additional term with free parameters}\label{ExtraCont}
As the last part of the solution, let us take the operator \cite{Smolyakov:2021vlo}
\begin{equation}\label{phiEX}
\hat\phi_{\textrm{ex}}(t,\vec x)=\frac{1}{\sqrt{L^{3}}}\left(\sum\limits_{j\neq 0}\left(\frac{1}{2}-\frac{i\omega t}{\hbar}\right)\epsilon_{j}\hat a_{j}^{\dagger}\hat a_{j}^{}-\sum\limits_{j\neq 0}\frac{\epsilon_{j}}{2q}e^{\frac{i}{\hbar}\vec k_{j}\vec x}\left(e^{-\frac{i}{\hbar}\gamma_{j}t}c_{j}\hat P_{j}-e^{\frac{i}{\hbar}\gamma_{j}t}d_{j}\hat P_{-j}^{\dagger}\right)\right)
\end{equation}
with $\hat P_{j}=\left(\hat a_{0}^{}-\hat a_{0}^{\dagger}\right)\hat a_{j}$. Here $\epsilon_{-j}=\epsilon_{j}$ are arbitrary dimensionless parameters. Since the operator $\hat\phi_{\textrm{ex}}(t,\vec x)$ is a solution of the homogeneous part of Eq.~\eqref{eqnlcquant2}, it can always be added to the solution for $\hat\phi(t,\vec x)$ obtained in the previous sections. It was shown in \cite{Smolyakov:2021vlo} that the necessary commutation relations \eqref{CCR1} and \eqref{CCR2} are satisfied for \eqref{phiEX}. As we will see in the next section, operator \eqref{phiEX} introduces free parameters into the resulting effective theory.

\section{Operators $\hat N_{p}$, $\hat E_{p}$ and $\hat{\vec{P_{p}}}$}
Up to the terms $\sim\beta$, solution for the operator $\hat\Psi(t,\vec x)$ has the form
\begin{align}\nonumber
\hat\Psi(t,\vec x)&=e^{-\frac{i}{\hbar}\omega t}\left(\rule{0cm}{1cm}\right.\sqrt{\frac{\omega}{g}}+\hat\varphi(t,\vec x)+\beta\Biggl(\hat\phi_{\textrm{no}}(t)+\hat\phi_{\times}(t,\vec x)+
\hat\phi_{\textrm{t}}(t)\\\label{solfull}&+\frac{1}{2}\underset{\substack{j\neq 0, l\neq 0\\j\neq-l}}{\sum\sum}\hat\phi_{j,l}^{\scriptscriptstyle+}(t,\vec x)+\frac{1}{2}\underset{\substack{j\neq 0, l\neq 0\\j\neq l}}{\sum\sum}\hat\phi_{j,l}^{\scriptscriptstyle-}(t,\vec x)+\hat\phi_{\textrm{ex}}(t,\vec x)\Biggr)\left.\rule{0cm}{1cm}\right).
\end{align}
As was checked above, this solution satisfies the necessary commutation relations \eqref{CCR1} and \eqref{CCR2} (i.e., the canonical commutation relations \eqref{commrel} up to the terms $\sim\beta$). Substituting \eqref{solfull} into \eqref{NpquadrQNL}, \eqref{EpquadrQNL} and into the momentum operator, which has the standard definition
\begin{equation}
\hat{\vec{P}}=-i\hbar\int d^{3}x\hat\Psi^{\dagger}(t,\vec x)\nabla\hat\Psi(t,\vec x),
\end{equation}
using the relation $\int d^{3}x\,e^{\pm\frac{i}{\hbar}\vec k_{j}\vec x}=0$, which is valid for any $\vec k_{j}\neq 0$, and taking $q=\sqrt{\frac{L^{3}}{\omega g}}\,\tilde q$, where $\tilde q$ is an arbitrary constant with dimension of $\omega$ (recall that the parameter $q$ originates from \eqref{quantmode0} and \eqref{quantmode}), after straightforward calculations one gets
\begin{align}
\label{Npfinal0}
\hat N_{p}&=\frac{1}{2}\left(1+\sum\limits_{j\neq 0}\tilde v_{j}\left(1-\frac{{\vec k_{j}}^{2}}{2m\gamma_{j}}\right)\right)+\frac{dN_{0}}{d\omega}\,\tilde q\left(\hat a_{0}^{}+\hat a_{0}^{\dagger}\right)+\sum\limits_{j\neq 0}\left(\epsilon_{j}-\frac{\tilde v_{j}{\vec k_{j}}^{2}}{2m\gamma_{j}}\right)\hat a_{j}^{\dagger}\hat a_{j}^{},\\
\nonumber
\hat E_{p}&=\frac{1}{2}\sum\limits_{j\neq 0}\frac{{\vec k_{j}}^{2}}{2m\gamma_{j}}\left(\frac{{\vec k_{j}}^{2}}{2m}+\omega\tilde v_{j}-\gamma_{j}\right)+
\frac{dE_{0}}{d\omega}\,\tilde q\left(\hat a_{0}^{}+\hat a_{0}^{\dagger}\right)+\frac{1}{2}\frac{d^{2}E_{0}}{d\omega^{2}}\,{\tilde q}^{2}\left(\hat a_{0}^{}+\hat a_{0}^{\dagger}\right)^{2}\\\label{Epfinal0}
&+\omega\sum\limits_{j\neq 0}\left(\epsilon_{j}-\frac{\tilde v_{j}{\vec k_{j}}^{2}}{2m\gamma_{j}}\right)\hat a_{j}^{\dagger}\hat a_{j}^{}+\sum\limits_{j\neq 0}\gamma_{j}\hat a_{j}^{\dagger}\hat a_{j}^{},\\\label{Ppfinal0}
\hat{\vec{P_{p}}}&=\sum\limits_{j\neq 0}\vec k_{j}\hat a_{j}^{\dagger}\hat a_{j}^{}.
\end{align}
In the latter formulas, the operator $\hat a_{j}^{\dagger}\hat a_{j}^{}$ corresponds to the number of quasi-particles with momentum $\vec k_{j}$, whereas the $c$-number terms in \eqref{Npfinal0} and \eqref{Epfinal0} are the consequence of permutations of the creation and annihilation operators. As in \cite{Smolyakov:2021vlo}, even for arbitrary $\epsilon_{j}$ the operators $\hat N_{p}$, $\hat E_{p}$ do not contain any nondiagonal terms like those in \cite{Bogolyubov}.

Choosing
\begin{equation}\label{epsilondef}
\epsilon_{j}=\frac{\tilde v_{j}{\vec k_{j}}^{2}}{2m\gamma_{j}}
\end{equation}
in \eqref{phiEX} (recall that $\epsilon_{j}$ are free parameters of the solution) and dropping the $c$-number terms that are not relevant for our analysis, for the operators in \eqref{Npfinal0}--\eqref{Ppfinal0} we obtain
\begin{align}\label{Npfinal}
&\hat N_{p}=\frac{dN_{0}}{d\omega}\,\tilde q\left(\hat a_{0}^{}+\hat a_{0}^{\dagger}\right),\\\label{Epfinal}
&\hat E_{p}=\frac{dE_{0}}{d\omega}\,\tilde q\left(\hat a_{0}^{}+\hat a_{0}^{\dagger}\right)+\frac{1}{2}\frac{d^{2}E_{0}}{d\omega^{2}}\,{\tilde q}^{2}\left(\hat a_{0}^{}+\hat a_{0}^{\dagger}\right)^{2}+\sum\limits_{j\neq 0}\gamma_{j}\hat a_{j}^{\dagger}\hat a_{j}^{},\\\label{Ppfinal}
&\hat{\vec{P_{p}}}=\sum\limits_{j\neq 0}\vec k_{j}\hat a_{j}^{\dagger}\hat a_{j}^{}.
\end{align}
As was explained in \cite{Smolyakov:2021vlo}, the terms with the operator $\hat a_{0}+\hat a_{0}^{\dagger}$ are expected for the nonoscillation mode corresponding to a change of $\omega$ in the background solution \eqref{backgrsol0}.\footnote{In full analogy with the result of \cite{Smolyakov:2021vlo}, the term $\sim\left(\hat a_{0}^{}+\hat a_{0}^{\dagger}\right)^{2}$ is absent in \eqref{Npfinal0} because $\frac{d^{2}N_{0}}{d\omega^{2}}=0$ (see \eqref{backgrEN}) for the background solution \eqref{backgrsol0}.} Exactly as in \cite{Smolyakov:2021vlo}, formula \eqref{Npfinal} implies that free quasi-particles do not change the particle number of the system --- it is the term with the operator $\hat a_{0}^{}+\hat a_{0}^{\dagger}$ that is responsible for adding particles to the system (more precisely, to the condensate) or removing them from the system (for a detailed discussion, see \cite{Smolyakov:2021vlo}). However, as expected, free quasi-particles can change the energy of the system (which is $\gamma_{j}$ for a quasi-particle, see \eqref{Epfinal}), and its momentum (which is $\vec k_{j}$ for a quasi-particle, see \eqref{Ppfinal}). In is interesting to note that the second term in \eqref{Epfinal} is similar to the term with the ``momentum'' operator of the condensate, which was introduced in \cite{LY}, in the Hamiltonian derived in \cite{LY}.

\section{Conclusion}
It is shown that the use of additional nonoscillation modes in the linear approximation together with the first nonlinear correction provides a natural solution to the problem of nonconserved particle number in the case of arbitrary two-body interaction potential as well. This happens in the same way as in the special case of simplified two-body interaction potential \eqref{interactpotent} \cite{Smolyakov:2021vlo}.\footnote{One can check that all formulas presented in this paper can be reduced to those of \cite{Smolyakov:2021vlo} by taking $v_{j}=v_{0}=g$ (i.e., $\tilde v_{j}=1$) for all $j$.} Within the proposed approach, which only slightly modifies the original method of Bogolyubov \cite{Bogolyubov}, the particle number is conserved automatically and there is no need for additional methods of the particle number conservation (like those that rely on modifications of the resulting effective Hamiltonian). An important point is that both parts of the approach --- the nonoscillation modes and the first nonlinear correction, are necessary to get a correct result.

It is necessary to stress that since
\begin{equation}
\hat\Psi(0,\vec x)\not\equiv\Psi_{0}(0,\vec x)+\hat\varphi_{\textrm{o}}^{}(0,\vec x),
\end{equation}
where $\hat\varphi_{\textrm{o}}^{}(t,\vec x)$ is defined by \eqref{linsol}, solution \eqref{solfull} cannot be considered just as an approximation of a solution of Eq.~\eqref{eqmotgeneral} with the usual initial condition $\Psi_{0}(0,\vec x)+\hat\varphi_{\textrm{o}}^{}(0,\vec x)$, see the corresponding discussion in \cite{Smolyakov:2021vlo}. Thus, it is the consideration of a different solution of the nonlinear integro-differential Eq.~\eqref{eqmotgeneral} that results in cancellation of the unphysical nondiagonal terms arising when one takes the standard solution. Certainly, solution \eqref{solfull} (as well as the original solution $\Psi_{0}(t,\vec x)+\hat\varphi_{\textrm{o}}^{}(t,\vec x)$ of \cite{Bogolyubov}) is not exact\footnote{Meanwhile, it provides a more accurate fulfillment of the canonical commutation relations than the original solution of \cite{Bogolyubov}, see \cite{Smolyakov:2021vlo}.} --- in order to get a more accurate solution, one should find at least several more terms in expansion \eqref{seriesPhiinf}. However, even calculation of the second nonlinear correction $\hat\phi_{2}(t,\vec x)$ seems to be a much more difficult task than calculation of the first nonlinear correction $\hat\phi_{1}(t,\vec x)$ presented in this paper. This problem, as well as calculation of the interaction Hamiltonian corresponding to the new solution $\hat\Psi(t,\vec x)$ defined by \eqref{solfull}, call for further investigation.

\section*{Appendix}
Let us consider equality \eqref{lasteqcoeff}. Using \eqref{L-def} and definitions \eqref{qdef3}, \eqref{qdef4}, the expression $L_{a+b,b}^{\scriptscriptstyle-}c_{b}-L_{a+b,a}^{\scriptscriptstyle-}c_{a}$ can be brought to the form
\begin{align}\nonumber
&-\frac{ic_{b}}{2\hbar\gamma_{a}}\Biggl(\Bigl(\sqrt{\gamma_{a}^{2}+(\omega\tilde v_{a})^{2}}+\gamma_{a}\Bigr)\Bigl((\tilde v_{a+b}+\tilde v_{a})d_{a+b}d_{b}+(\tilde v_{a}+\tilde v_{b})c_{a+b}c_{b}-(\tilde v_{a+b}+\tilde v_{b})c_{a+b}d_{b}\Bigr)\\\nonumber&
-\omega\tilde v_{a}\Bigl((\tilde v_{a}+\tilde v_{b})d_{a+b}d_{b}+(\tilde v_{a+b}+\tilde v_{a})c_{a+b}c_{b}-(\tilde v_{a+b}+\tilde v_{b})d_{a+b}c_{b}\Bigr)\Biggr)\\\nonumber
&+\frac{ic_{a}}{2\hbar\gamma_{b}}\Biggl(\Bigl(\sqrt{\gamma_{b}^{2}+(\omega\tilde v_{b})^{2}}+\gamma_{b}\Bigr)\Bigl((\tilde v_{a+b}+\tilde v_{b})d_{a+b}d_{a}+(\tilde v_{a}+\tilde v_{b})c_{a+b}c_{a}-(\tilde v_{a+b}+\tilde v_{a})c_{a+b}d_{a}\Bigr)\\\label{AppDexpr}&
-\omega\tilde v_{b}\Bigl((\tilde v_{a}+\tilde v_{b})d_{a+b}d_{a}+(\tilde v_{a+b}+\tilde v_{b})c_{a+b}c_{a}-(\tilde v_{a+b}+\tilde v_{a})d_{a+b}c_{a}\Bigr)\Biggr).
\end{align}
With the help of relations
\begin{equation}\label{AppDrels}
c_{j}d_{j}=\frac{\omega\tilde v_{j}}{2\gamma_{j}},\qquad c_{j}^{2}=\frac{(\omega\tilde v_{j})^{2}}{2\gamma_{j}\left(\sqrt{\gamma_{j}^{2}+(\omega\tilde v_{j})^{2}}-\gamma_{j}\right)}
=\frac{\sqrt{\gamma_{j}^{2}+(\omega\tilde v_{j})^{2}}+\gamma_{j}}{2\gamma_{j}},
\end{equation}
which follow from \eqref{cddef}, one can get for \eqref{AppDexpr}
\begingroup
\allowdisplaybreaks
\begin{align}\nonumber
&-\frac{i}{4\hbar\gamma_{a}\gamma_{b}}\Biggl[\left(\sqrt{\gamma_{a}^{2}+(\omega\tilde v_{a})^{2}}+\gamma_{a}\right)\Biggl(
(\tilde v_{a+b}+\tilde v_{a})d_{a+b}\omega\tilde v_{b}+(\tilde v_{a}+\tilde v_{b})c_{a+b}\left(\sqrt{\gamma_{b}^{2}+(\omega\tilde v_{b})^{2}}+\gamma_{b}\right)\\\nonumber&-(\tilde v_{a+b}+\tilde v_{b})c_{a+b}\omega\tilde v_{b}\Biggr)-\omega\tilde v_{a}\Biggl((\tilde v_{a}+\tilde v_{b})d_{a+b}\omega\tilde v_{b}+(\tilde v_{a+b}+\tilde v_{a})c_{a+b}\left(\sqrt{\gamma_{b}^{2}+(\omega\tilde v_{b})^{2}}+\gamma_{b}\right)\\\nonumber&
-(\tilde v_{a+b}+\tilde v_{b})d_{a+b}\left(\sqrt{\gamma_{b}^{2}+(\omega\tilde v_{b})^{2}}+\gamma_{b}\right)\Biggr)\\\nonumber
&-\left(\sqrt{\gamma_{b}^{2}+(\omega\tilde v_{b})^{2}}+\gamma_{b}\right)\Biggl(
(\tilde v_{a+b}+\tilde v_{b})d_{a+b}\omega\tilde v_{a}+(\tilde v_{a}+\tilde v_{b})c_{a+b}\left(\sqrt{\gamma_{a}^{2}+(\omega\tilde v_{a})^{2}}+\gamma_{a}\right)\\\nonumber&-(\tilde v_{a+b}+\tilde v_{a})c_{a+b}\omega\tilde v_{a}\Biggr)+\omega\tilde v_{b}\Biggl((\tilde v_{a}+\tilde v_{b})d_{a+b}\omega\tilde v_{a}+(\tilde v_{a+b}+\tilde v_{b})c_{a+b}\left(\sqrt{\gamma_{a}^{2}+(\omega\tilde v_{a})^{2}}+\gamma_{a}\right)\\&
-(\tilde v_{a+b}+\tilde v_{a})d_{a+b}\left(\sqrt{\gamma_{a}^{2}+(\omega\tilde v_{a})^{2}}+\gamma_{a}\right)\Biggr)\Biggr].
\end{align}
\endgroup
Combining the similar terms, it is easy to see that this expression is equal to zero identically. Thus, equality \eqref{lasteqcoeff} is satisfied.

Now let us consider equality \eqref{lastbut1eqcoeff}. Using \eqref{L+def}, \eqref{L-def} and definitions \eqref{qdef1}--\eqref{qdef4}, the expression for $L_{a,b}^{\scriptscriptstyle+}c_{b}-L_{a+b,a}^{\scriptscriptstyle-}c_{a+b}$ can be brought to the form
\begin{align}\nonumber
&-\frac{ic_{b}}{2\hbar(\gamma_{a}+\gamma_{b})}\Biggl(\left(\sqrt{\left(\gamma_{a}+\gamma_{b}\right)^{2}+(\omega\tilde v_{a+b})^{2}}+\gamma_{a}+\gamma_{b}\right)\Bigl((\tilde v_{a}+\tilde v_{b})c_{a}c_{b}-(\tilde v_{a+b}+\tilde v_{b})c_{a}d_{b}\\\nonumber&-(\tilde v_{a+b}+\tilde v_{a})d_{a}c_{b}\Bigr)
-\omega\tilde v_{a+b}\Bigl((\tilde v_{a}+\tilde v_{b})d_{a}d_{b}-(\tilde v_{a+b}+\tilde v_{a})c_{a}d_{b}-(\tilde v_{a+b}+\tilde v_{b})d_{a}c_{b}\Bigr)\Biggr)\\\nonumber
&+\frac{ic_{a+b}}{2\hbar\gamma_{b}}\Biggl(\left(\sqrt{\gamma_{b}^{2}+(\omega\tilde v_{b})^{2}}+\gamma_{b}\right)\Bigl((\tilde v_{a+b}+\tilde v_{b})d_{a+b}d_{a}+(\tilde v_{a}+\tilde v_{b})c_{a+b}c_{a}-(\tilde v_{a+b}+\tilde v_{a})c_{a+b}d_{a}\Bigr)\\\label{AppDexpr2}&
-\omega\tilde v_{b}\Bigl((\tilde v_{a}+\tilde v_{b})d_{a+b}d_{a}+(\tilde v_{a+b}+\tilde v_{b})c_{a+b}c_{a}-(\tilde v_{a+b}+\tilde v_{a})d_{a+b}c_{a}\Bigr)\Biggr).
\end{align}
With the help of relations \eqref{AppDrels}, one can get for \eqref{AppDexpr2}
\begin{align}\nonumber
&-\frac{i}{4\hbar(\gamma_{a}+\gamma_{b})\gamma_{b}}\Biggl[\left(\sqrt{\left(\gamma_{a}+\gamma_{b}\right)^{2}+(\omega\tilde v_{a+b})^{2}}+\gamma_{a}+\gamma_{b}\right)\\\nonumber&\times
\Biggl(\left(\sqrt{\gamma_{b}^{2}+(\omega\tilde v_{b})^{2}}+\gamma_{b}\right)\Bigl((\tilde v_{a}+\tilde v_{b})c_{a}-(\tilde v_{a+b}+\tilde v_{a})d_{a}\Bigr)
-(\tilde v_{a+b}+\tilde v_{b})c_{a}\omega\tilde v_{b}\Biggr)\\\nonumber
&-\omega\tilde v_{a+b}\left((\tilde v_{a}+\tilde v_{b})d_{a}\omega\tilde v_{b}-(\tilde v_{a+b}+\tilde v_{a})c_{a}\omega\tilde v_{b}-
\left(\sqrt{\gamma_{b}^{2}+(\omega\tilde v_{b})^{2}}+\gamma_{b}\right)(\tilde v_{a+b}+\tilde v_{b})d_{a}\right)\\\nonumber
&-\left(\sqrt{\gamma_{b}^{2}+(\omega\tilde v_{b})^{2}}+\gamma_{b}\right)
\Biggl((\tilde v_{a+b}+\tilde v_{b})d_{a}\omega\tilde v_{a+b}+\left(\sqrt{\left(\gamma_{a}+\gamma_{b}\right)^{2}+(\omega\tilde v_{a+b})^{2}}+\gamma_{a}+\gamma_{b}\right)\\\nonumber
&\times\Bigl((\tilde v_{a}+\tilde v_{b})c_{a}-(\tilde v_{a+b}+\tilde v_{a})d_{a}\Bigr)\Biggr)+\omega\tilde v_{b}\Biggl((\tilde v_{a}+\tilde v_{b})d_{a}\omega\tilde v_{a+b}\\&+
\left(\sqrt{\left(\gamma_{a}+\gamma_{b}\right)^{2}+(\omega\tilde v_{a+b})^{2}}+\gamma_{a}+\gamma_{b}\right)(\tilde v_{a+b}+\tilde v_{b})c_{a}
-(\tilde v_{a+b}+\tilde v_{a})c_{a}\omega\tilde v_{a+b}\Biggr)\Biggr].
\end{align}
Combining the similar terms, it is easy to see that this expression is equal to zero identically. Thus, equality \eqref{lastbut1eqcoeff} is also satisfied.

\section*{Supplementary material}
Supplementary material contains the file for the computer algebra system Maxima with wxMaxima interface \cite{Maxima}, which was used to compute some coefficients in the commutators presented in this paper, and its .pdf version allowing one to look at the result of the computations without installing the program package. Maxima 5.44.0 (with wxMaxima 20.06.6) was used for the computations.

\end{document}